\documentclass[a4paper,12pt]{article}

\usepackage{orcidlink,thumbpdf,lmodern}
\usepackage{amsmath}
\usepackage{amsthm}
\usepackage[ruled,vlined]{algorithm2e}
\usepackage{amsfonts}
\usepackage[hypcap=false]{caption}
\usepackage{listings}
\usepackage{lscape}
\usepackage{natbib}
\usepackage{graphicx}
\usepackage{authblk}
\usepackage{geometry} 
\newcommand{\vect}[1]{\ensuremath{\boldsymbol{#1}}}
\newcommand{\mat}[1]{\ensuremath{\boldsymbol{#1}}}

\newtheorem{example}{Example}[section]

\begin{document}

\title{Goodness-of-Fit and Clustering of Spherical Data: The \texttt{QuadratiK} package in R and Python}

\author[1]{Giovanni Saraceno\thanks{\href{mailto:gsaracen@buffalo.edu}{gsaracen@buffalo.edu}}\orcidlink{0000-0002-1753-2367}}
\author[1]{Marianthi Markatou\thanks{Corresponding author: \href{mailto:markatou@buffalo.edu}{markatou@buffalo.edu}}\orcidlink{0000-0002-1453-8229}}
\author[1,2]{Raktim Mukhopadhyay\thanks{\href{mailto:raktimmu@buffalo.edu}{raktimmu@buffalo.edu}}\orcidlink{0009-0007-8972-6682}}
\author[3]{Mojgan Golzy\thanks{\href{mailto:golzym@health.missouri.edu}{golzym@health.missouri.edu}}\orcidlink{0000-0002-2759-3120}}

\affil[1]{Department of Biostatistics, University at Buffalo, NY, USA}
\affil[2]{CDSE Program\\University at Buffalo}
\affil[3]{Biomedical Informatics, Biostatistics and Medical Epidemiology (BBME), University of Missouri}
\date{}

\maketitle

\begin{abstract}
	We introduce the \texttt{QuadratiK} package that incorporates innovative data analysis methodologies. The presented software, implemented in both R and Python, offers a comprehensive set of goodness-of-fit tests and clustering techniques using kernel-based quadratic distances, thereby bridging the gap between the statistical and machine learning literatures. Our software implements one, two and k-sample tests for goodness of fit, providing  an efficient and mathematically sound way to assess the fit of probability distributions. Expanded capabilities of our software include supporting tests for uniformity on the $d$-dimensional Sphere based on Poisson kernel densities. Particularly noteworthy is the incorporation of a unique clustering algorithm specifically tailored for spherical data that leverages a mixture of Poisson kernel-based densities on the sphere. Alongside this, our software includes additional graphical functions, aiding the users in validating, as well as visualizing and representing clustering results. This enhances interpretability and usability of the analysis. 
	In summary, our R and Python packages serve as a powerful suite of tools, offering researchers and practitioners the means to delve deeper into their data, draw robust inference, and conduct potentially impactful analyses and inference across a wide array of disciplines. 

    \smallskip 
    
	
\end{abstract}

\maketitle

\section{Introduction}
\label{sec:introduction}

\texttt{QuadratiK} is an open-source, comprehensive package, written in \texttt{R} and \texttt{Python}, of multivariate analysis and inference methods that are based on quadratic distances \cite{markatou2021}. \texttt{QuadratiK} allows users to perform goodness-of-fit tests using very large, multidimensional data sets. These data sets are met in many scientific fields from biology, to medicine, to agriculture and public health. The UCI Machine Learning Repository (https://archieve.ics.uci.edu) provides many such datasets. The package offers the user the capability to perform one, two, and $k$-sample goodness-of-fit tests, as well as to perform data clustering using the method developed by \cite{golzy2020}.

The problem of Goodness-of-Fit (GoF) is crucial in research. GoF tests constitute a classical tool to decide the comparability of data with an assumed probability model (distribution). Comparing two or more samples allows researchers to examine differences or similarities between groups or conditions. This can help identify factors that may contribute to certain outcomes or behaviours.  Furthermore, such comparisons facilitate the validation of findings and ensure the generalizability of research results, thus enhance the robustness and reliability of research findings.
The standard way of constructing a GoF test procedure involves the computation of some distance-like functional between the null distribution and the given observations, where the null hypothesis is rejected if the obtained distance is greater than a critical value. 

The statistical literature includes many GoF tests, and software written in both, \texttt{R} and \texttt{Python}, encodes many of the proposed methods. Our interest centers on a family of kernel-based quadratic distances (KBQDs), a set of methods that are fundamentally different with all other methods that appear in the literature. 

Quadratic distances are central to the study of GoF and important statistics that belong to this family are the Pearson's chi-squared and Cram\'er-von Mises statistics. \cite{Lindsay2008} introduce a unified framework for the study of quadratic distance measures, and \cite{Lindsay2014} study the power properties of kernel-based quadratic distance goodness-of-fit tests introducing the midpower analysis and the concept of degrees of freedom for the kernel function. The authors, then, investigate the power properties when testing for multivariate normality by using the normal kernel. \cite{markatou2021} offer a comprehensive review of the quadratic distance-based methodologies, while \cite{Markatou2017} offer interesting interpretations of distances and discuss their role in robustness. Additionally, \cite{Markatou2023} introduce a unified framework for constructing two- and $k$-sample hypothesis tests using kernel-based quadratic distances. The kernels we use are diffusion kernels \citep{Ding2023}, that is, probability distributions that depend on a tuning parameter and satisfy the convolution property. \cite{Ding2023} propose the Poisson kernel-based tests for uniformity on the $d$-dimensional sphere. Indeed, the Poisson kernel appears to be the natural generalization of the normal distribution to the sample space $\mathcal{S}^{d-1}$, and through a normalizing constant, the Poisson kernel can be considered as a density on the sphere. Applications of the Poisson kernel-based densities to clustering can be found in \cite{golzy2020} and \cite{Golzy2016}.  

GoF methods encoded in \texttt{R} and \texttt{Python} deal extensively with univariate data. Relatively less attention has been given to multivariate two- and $k$-sample tests in multiple-dimensional settings. Some of the available multivariate methods for the two-sample case utilize graph-based multivariate ranking strategies to extend univariate two-sample GoF tests in high dimensional settings. One such approach was introduced by \cite{Friedman1979}, who employed the minimal spanning tree (MST) as a  multivariate ordered list. These authors extend the univariate run-based test introduced by \cite{Wald1940}, the univariate two-sample Kolmogorov-Smirnov test and a modified Kolmogorov-Smirnov test against scale alternatives. 
Similarly, \cite{Biswas2014a} adopts the shortest Hamiltonian path instead of MST as ranking strategy for multivariate observations. \cite{Chen2017} presents a test statistic based on a similarity graph. In addition to graph-based methods, general distance measures have been widely adopted to handle the multivariate two-sample GoF testing problem. \cite{Rosembaum2005} introduced a two-sample test based on interpoint distances. 
From the community of Machine Learning, \cite{Gretton2012} construct the test statistic called Maximum Mean Discrepancy (MMD), utilizing the properties of the kernel mean embedding. 
The operating characteristics of several of the tests mentioned above, are studied in \cite{Chen2020}.

The two-sample problem can be extended to the $k$-sample testing problem, where more than two groups are compared.
The Anderson-Darling test, initially introduced for the case of testing equality of two distributions, was extended by \cite{Scholz1987}. 
An alternative to nonparametric test construction is given by the general framework of permutation tests. 
\citep{coin} provides an implementation of such tests tailored against location and scale alternatives, and for survival distributions. \cite{Rizzo2010} propose a nonparametric extension of the standard analysis of variance procedure, called Distance Components (DISCO), using the Energy statistics. 
Kernel-based methods have gained popularity also for the $k$-sample problem due to their ability to handle high-dimensional data.  \cite{balogoun2022} propose a $k$-sample test for functional data by introducing a generalization of the MMD. Recently, \cite{panda2019nonpar} highlighted the connection between $k$-sample testing and independence testing problems showing that independence tests, such as the MMD and the Energy statistics, can be used for consistent $k$-sample testing. 

Methods to test for uniformity on the sphere are also provided in the literature. \texttt{QuadratiK} includes a test for uniformity for spherical observations that is based on the Poisson kernel density. 

\textit{Contributions.} 
\texttt{QuadratiK} offers a set of multivariate analysis procedures that are based on the special class diffusion kernels. The implementation methods are practical and accessible, and they can handle high dimensional data sets. 
In particular, the test procedures employ optimized \text{C++} versions which speed up the calculations. 
The implementation of the quadratic distance tests involves the following steps. First, the kernels need to be centered. Centering of the kernel is done both parametrically and nonparametrically. 
Second, the selection of the tuning parameter of the kernel and the computation of the critical value, incorporating various sampling algorithms, are a very important aspect for computing the tests.
Indeed, the software offers the possibility to select the optimal value of the tuning parameter according to the midpower analysis. 
By employing the use of parallel computing, the algorithm exhibits a feasible computational time. To facilitate the interpretability and usability of the clustering results, the package includes graphical functions, which aid the users in validating and visualizing clustering results.
The presented software is implemented as \text{R} and \text{Python} packages in order to enhance their adoption into preexisting workflows by users from statistics and computer science communities, thus enhancing the dissemination and use of these tools.   

\textit{Outline}. The rest of the article is organized as follows. Section \ref{sec:KBQD} provides the necessary background on the kernel-based quadratic distances and introduces the structure of the \texttt{QuadratiK} package. Section \ref{sec:tests} details the construction of the KBQD tests. In the corresponding subsections, the test statistics for normality, the non-parametric two-sample and $k$-sample tests are derived using the normal kernel. These methods are introduced together with synthetic examples and two real data examples with codes in \texttt{R}. This section also includes the algorithm proposed for selecting the bandwidth parameter $h$ of the normal kernel based on the power analysis of the test. Finally, subsection \ref{subsec:test_uniform} introduces the Poisson kernel and the derived test statistics for testing uniformity on the sphere along with code illustrations. 
The clustering algorithm appropriate for use with data on the sphere \citep{golzy2020} provided in the software is described in Section \ref{sec:clustering}, and its usage, together with the additional graphical functions, is shown through a real data application. A summary is provided in Section \ref{sec:conclusion}. 

\section{Kernel-based Quadratic Distances}
\label{sec:KBQD}

The \texttt{QuadratiK} package provides a comprehensive set of tools for handling statistical tests and clustering operations based on kernel-based quadratic distances. 

\begin{figure}[htb]
    \centering
    \includegraphics[scale=0.6]{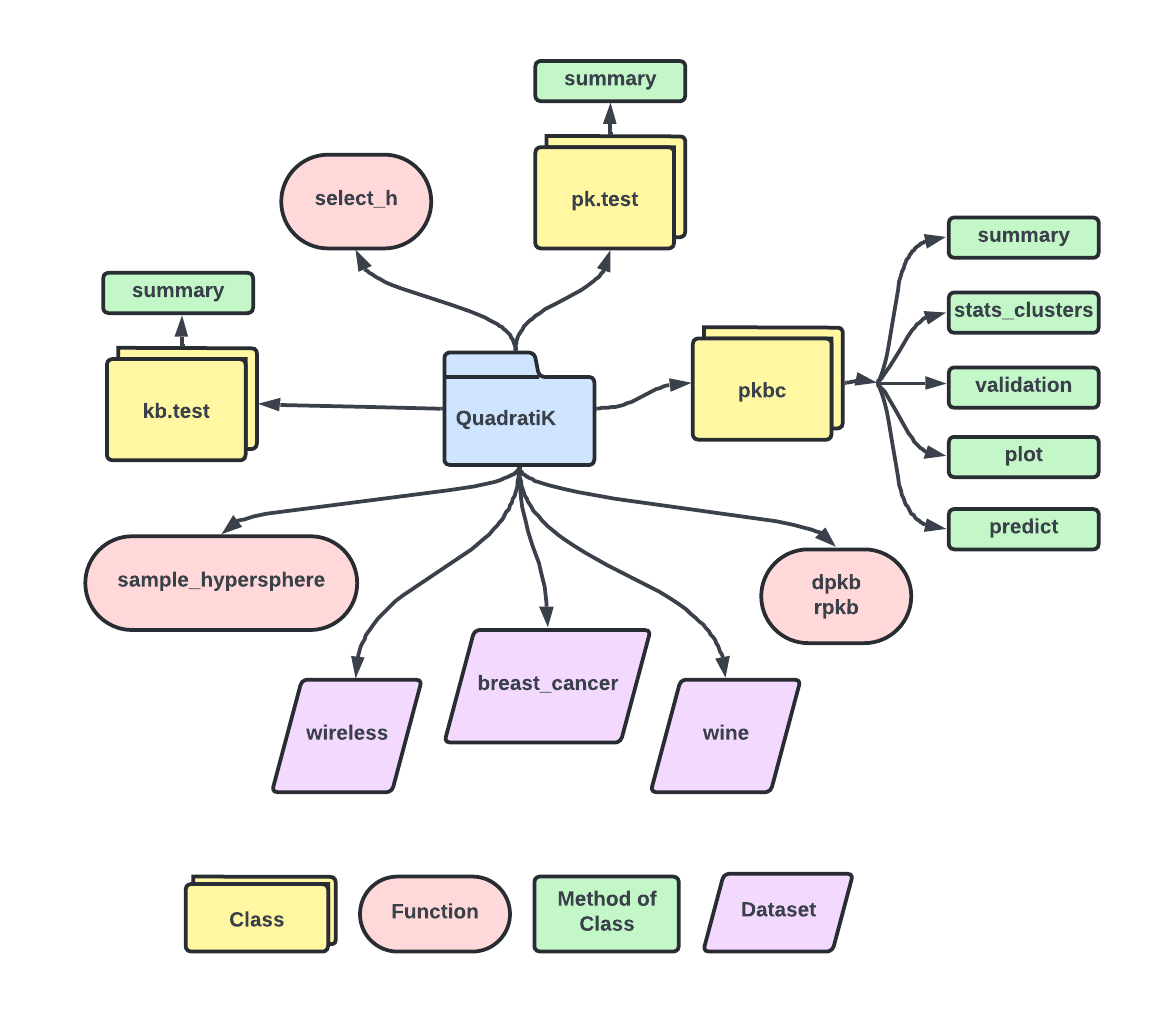}
    \caption{The classes along with their corresponding methods, and functions available in the current version of \texttt{QuadratiK} in \texttt{R} are shown here. The folder (blue) in the center represents the \texttt{QuadratiK} package. The double rectangle (yellow) depicts the classes, the rectangles (green) denote the methods associated with these classes. The rounded rectangles (rose-colored) represent the functions, and the parallelograms (lavender-colored) show the datasets that are available in the package.}
    \label{fig:structure-R}
\end{figure}
The organization of the \texttt{R} package is showed in Figure \ref{fig:structure-R}. Its core functionality includes the function \texttt{kb.test}, a versatile function designed for performing kernel-based quadratic distance goodness-of-fit tests applicable to one-sample normality tests, as well as two-sample and k-sample comparisons, utilizing a Gaussian kernel. Furthermore, the function \texttt{select\_h} is provided for optional tuning parameter selection. Employing parallel processing techniques through \texttt{doParallel}, the package enhances the performance for computationally intensive tasks making the selection of optimal tuning parameter accessible. 
Additionally, the function \texttt{pk.test} can be used to perform the test for uniformity for spherical observations. The results of the functions \texttt{kb.test} and \texttt{pk.test} are returned as S4 classes, with the main information of the test as attributes, such as the computed test statistics, corresponding critical values, data and used tuning parameter. Summary descriptions of the respective tests can be accessed via the \texttt{print} and \texttt{summary} methods for the \texttt{kb.test} and \texttt{pk.test} class objects. For clustering, the package introduces the function \texttt{pkbc} which allows to enter a vector of possible values as number of clusters. An object of the S4 class \texttt{pkbc} is returned, containing other than the input information, a list with the clustering results for each provided value of number of clusters. Several methods for the \texttt{pkbc} class are available within \texttt{QuadratiK}, for facilitating the understanding of complex clustering dynamics and assisting in determining the optimal number of clusters. The \texttt{show} and \texttt{summary} methods furnish comprehensive summaries of the clustering results. The \texttt{stats\_clusters} method delivers descriptive statistics with respect to the clusters identified, while the \texttt{predict} method allows for predicting the membership of new data. Finally, the \texttt{plot} method is designed to support detailed graphical representations of clustering outcomes, including the elbow plot for identifying the number of clusters and a scatter plot for data point visualization. In addition, the function \texttt{pkbc\_validation} provides several statistics used in literature for addressing the choice of the number of clusters. Finally, the package includes the functions \texttt{dpkb}, which compute the density value, and \texttt{rpkb}, which generates random samples from the Poisson kernel based density. These functions are not discussed here, for further information please see the vignette \href{https://giovsaraceno.github.io/QuadratiK-package/articles/generate_rpkb.html}{https://giovsaraceno.github.io/QuadratiK-package/articles/generate\_rpkb.html}.

\begin{figure}[ht]
    \centering
    \includegraphics[width=\linewidth]{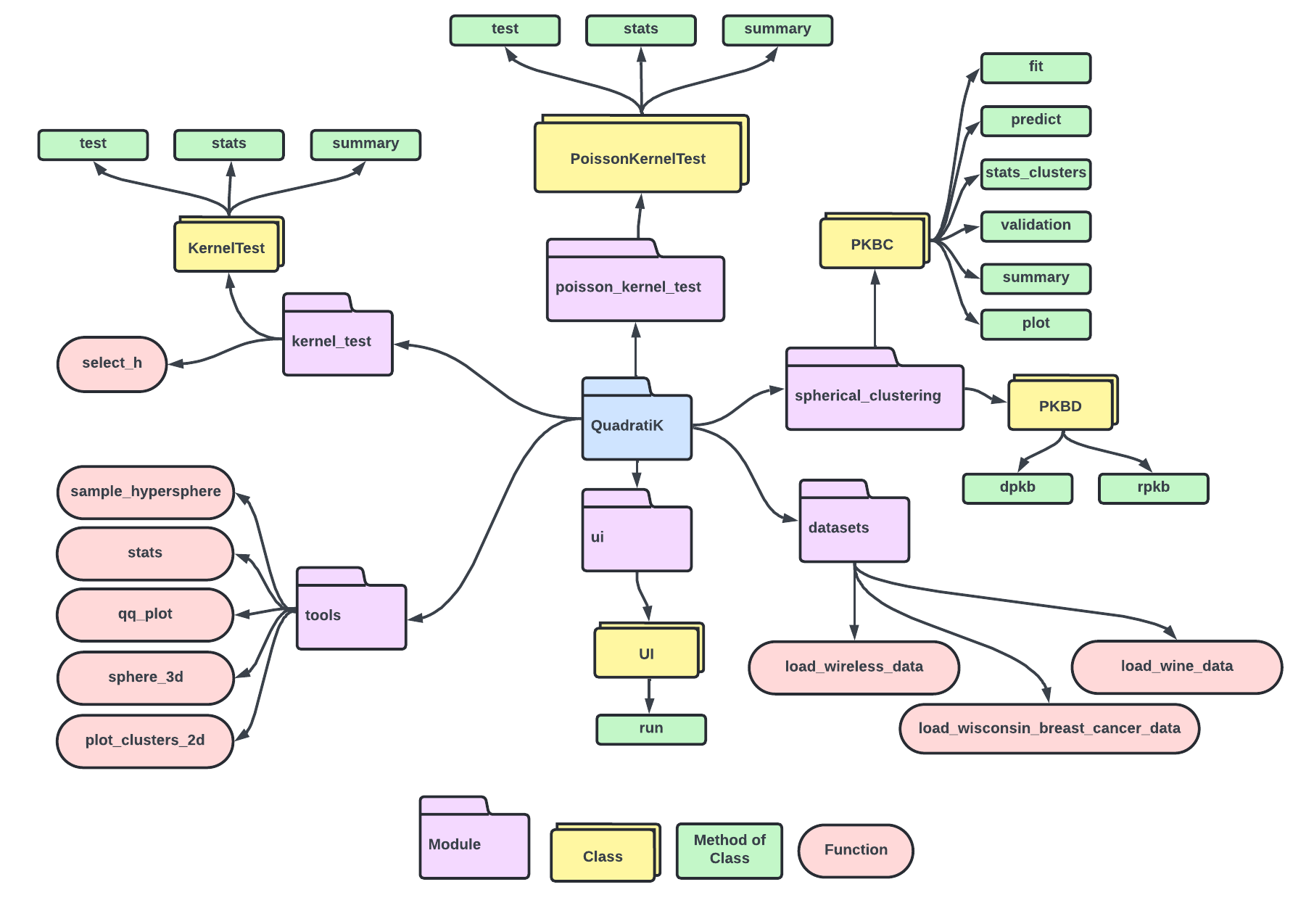}
    \caption{The main modules, along with their classes, methods, and functions, available in the current version of \texttt{QuadratiK} in \texttt{Python}. The folder (blue) in the center represents the \texttt{QuadratiK} package. The folders (lavender-colored) depict the various modules within the \texttt{QuadratiK}. The classes within these modules are depicted by double rectangles (yellow), the methods associated with these classes are shown as rectangles (green). Additionally, the package includes various functions, represented by rounded rectangles (rose-colored), which are included in different modules and offer various utilities.}
    \label{fig:structure-python}
\end{figure}
The structure of the \texttt{QuadratiK} package in \texttt{Python} is represented in Figure \ref{fig:structure-python}.
The Python package comprises modules designed for various GoF tests and clustering, along with a \texttt{tools} module. The \texttt{kernel\_test} module includes the \texttt{KernelTest} class, which is used to perform the normality, two-sample, and k-sample tests using a Gaussian kernel. The \texttt{KernelTest} class also supports methods such as \texttt{stats} and \texttt{summary} to enhance interaction with the provided functionalities. Additionally, the \texttt{kernel\_test} module contains the \texttt{select\_h} function for computing the optimal bandwidth parameter. The \texttt{poisson\_kernel\_test} module contains the \texttt{PoissonKernelTest} class for performing uniformity tests. Similar to the \texttt{R} implementation, this \texttt{Python} package provides essential test information as attributes, including computed test statistics, corresponding critical values, data, and tuning parameters for both the \texttt{KernelTest} and \texttt{PoissonKernelTest} classes. The \texttt{spherical\_clustering} module contains the \texttt{PKBC} class for performing Poisson kernel-based clustering. This module also includes methods to return descriptive statistics, elbow plots, and evaluation measures, aligning with the features available in the \texttt{R} implementation. The python implementation uses the \texttt{joblib} package to parallelize computations, improving efficiency for computing critical values for the GoF tests and determining the optimal bandwidth using the \texttt{select\_h} algorithm. The \texttt{tools} module offers common functions useful for descriptive statistics, visualizing data on a circle or sphere, generating QQ plots, and generating data on a hypersphere. Furthermore, the PKBD class contains methods \texttt{dpkb} and \texttt{rpkb} which can be used for computing the density and generating random samples from Poisson kernel based density respectively. The \texttt{Python} implementation includes an additional feature in the form of a user interface accessible through the \texttt{UI} class in the \texttt{ui} module, allowing users to access the methods implemented in the package without the need for programming. The \texttt{PKBD} and \texttt{UI} classes and corresponding methods are not discussed in this work, for more details please see the examples at \href{https://quadratik.readthedocs.io/en/latest/user_guide/gen_plot_rpkb.html}{https://quadratik.readthedocs.io/en/latest/user\_guide/gen\_plot\_rpkb.html}. 

In the following sections we describe in detail the \texttt{QuadratiK} functions. However, before we do so, we first introduce the theoretical background on the kernel-based quadratic distances.  
Given two probability distributions $F$ and $G$, the KBQD $d_K(F,G)$ is defined as 
\begin{equation}
	d_K(F,G) = \iint K_G(\vect{s},\vect{t}) d(F-G)(\vect{s}) d(F-G)(\vect{t}),
\end{equation}
where $K_G(\vect{s},\vect{t})$ denotes a non-negative definite kernel function that possibly depends on $G$. 
Uniqueness considerations of the value of distance associated with the kernel on which the distance is based dictate the use of centered kernels. The $G$-centered kernel, where $G$ is the probability distribution to be fitted to the data is defined as 
\begin{equation*}
	K_{\mathrm{cen}(G)}(\vect{s},\vect{t}) = K(\vect{s},\vect{t}) - K(\vect{s},G) - K(G,\vect{t}) + K(G,G),
\end{equation*}
 with $K(\vect{s},G) = \int K(\vect{s},\vect{t}) dG(\vect{t})$, $K(G,\vect{t}) = \int K(\vect{s},\vect{t}) dG(\vect{s})$, $K(G,G) = \iint K(\vect{s},\vect{t}) dG(\vect{s})dG(\vect{t})$. A cornerstone in constructing and using KBQDs is the choice of the kernel function. We concentrate on the special class of \textit{diffusion kernels}, introduced in \cite{Ding2023}. A family of symmetric kernels $K_t(\vect{s},\vect{t})$ with parameter $t\in (0, \infty)$ is called a diffusion kernel family with respect to a measure $\gamma$ if the following properties are satisfied. 
\begin{itemize}
\item The kernel is nonnegative, i.e. $K_t(\vect{s},\vect{t}) \ge 0$.
\item The kernel satisfies the diffusion equation 
\begin{equation}
K_{t_1 + t_2}(\vect{s},\vect{t}) = \int K_{t_1}(\vect{s},\vect{r}) K_{t_2}(\vect{r},\vect{t})dr.
\end{equation}
\item The kernel is a probability density under the measure $\gamma$ with respect to both arguments, that is 
\begin{equation*}
\int K_{t}(\vect{s},\vect{t})d\gamma(\vect{s}) = 1 \qquad \int K_{t}(\vect{s},\vect{t})d\gamma(\vect{t}) = 1. 
\end{equation*}
\end{itemize}
This class of kernels allows easy computation. Furthermore, the kernels are "tunable", since they depend on a tuning parameter $t$, connected with their degrees of freedom (DOF). The DOF of the kernel, in turn, drive the power characteristics of tests based on KBQDs (see \cite{Lindsay2008, Lindsay2014, Markatou2023} and \cite{Ding2023} for further information). 

\section{Multivariate KBQD tests}
\label{sec:tests}

In this section, we describe the main functions offered by the \texttt{QuadratiK} package which perform the multivariate KBQD test procedures. We illustrate the computational aspects of the kernel-based quadratic distance tests, including the choice of the kernel function, the selection of the corresponding tuning parameter, and the computation of the test statistics and corresponding critical values for assessing the hypothesis testing problem. We will show their usage through simulated data examples.  

\subsection{The function kb.test}

The \texttt{kb.test} function performs the kernel-based quadratic distance Goodness-of-Fit tests using the Gaussian kernel with tuning parameter $h$, that is the test for normality, two-sample test and $k$-sample test. For $\vect{s},\vect{t} \in \mathbb{R}^d$ and covariance matrix $\mat{\Sigma}_h$, depending on the tuning parameter $h$, it is given as 
$$ K_{{h}}(\vect{s}, \vect{t}) = (2 \pi)^{-d/2} \left(\det{\mat{\Sigma}_h}\right)^{-\frac{1}{2}}  \exp\left\{-\frac{1}{2}(\vect{s} - \vect{t})^\top \mat{\Sigma}_h^{-1}(\vect{s} - \vect{t})\right\}.$$
In our current implementation, we employ a covariance matrix $\mat{\Sigma}_h = h^2 \mat{I}_d$, where $\mat{I}_d$ represents the $d$-dimensional identity matrix. It is worth noting that more complex variance-covariance structures can be explored. For instance, one potential extension involves considering $\mat{\Sigma}_h$ as a diagonal matrix with diagonal elements $(h_1, h_2, \ldots, h_d)$. This approach introduces additional parameters, $h_i$, $i = 1, \ldots, d$, which allow for capturing different scales along each dimension of the data. Such flexibility can be valuable in scenarios where the underlying data exhibits varying degrees of correlation or dispersion across dimensions. While our current implementation focuses on the simpler $\mat{\Sigma}_h = h^2 \mat{I}_d$ case, the introduced extension does not appear to be straightforward and it is under consideration for future versions of our package, for potentially enhancing the modeling capabilities and adaptability.

The function \texttt{kb.test} takes the following arguments.
\begin{itemize}
    \item \texttt{x}: numeric matrix or vector of data values.
    \item \texttt{y}: NULL or numeric matrix or vector of data values or vector of group assignments.
    \item \texttt{h}: tuning parameter.
    \item \texttt{method}: sampling algorithm ("subsampling","bootstrap","permutation")
    \item \texttt{B}: the number of iterations to use for critical value estimation, B = 150 as default.
    \item \texttt{b}: subsample size in the subsampling algorithm.
    \item \texttt{Quantile}: the quantile to use for critical value estimation, 0.95 is the default value.
    \item \texttt{mu\_hat}: Mean vector for the reference distribution.
    \item \texttt{Sigma\_hat}: Covariance matrix of the reference distribution.
    \item \texttt{centeringType}: String indicating the method used for centering the normal kernel, "Param" or "Nonparam". (default: "Nonparam"). For $k$-sample tests only "Nonparam" is available.
    \item \texttt{alternative}: Family of alternatives chosen for selecting $h$, between \textit{location}, \textit{scale} and \textit{skewness} (only if $h$ is not provided). 
\end{itemize}
The specific test computed by the \texttt{kb.test} function depends on the input \texttt{y}. If \texttt{y} is not provided, the function computes the test for normality of the sample \texttt{x}. If \texttt{y} is a vector indicating group assignments, the function performs the $k$-sample test. Additionally, the special case of the two-sample problem can be assessed if \texttt{x} and \texttt{y} are the two samples to be compared.
Except for the data entry \texttt{x} (and \texttt{y}), and the bandwidth parameter \texttt{h}, the other arguments do not need to be specified. In such case, the implementation follows the default setting as specified in the list above. 
The function returns a class object with the results and summary of the performed test. 
In the case that the value of the tuning parameter $h$ is unknown, the provided function \texttt{select\_h} selects the "optimal" value of $h$ according to the mid-power analysis algorithm in \citep{Markatou2023}. The description and implementation of this algorithm are treated in section \ref{subsec:select-h}.  

In the following examples, we show the usage of the \texttt{kb.test} function in different contexts in which the test for normality, the two-sample test or the $k$-sample test, are needed. These examples present the codes in \texttt{R}, and for each of them the obtained results are displayed together with the needed computational time. Corresponding examples with Python code are provided as vignettes of the \texttt{Python} package \texttt{QuadratiK}, that can be found at the following link \href{https://quadratik.readthedocs.io/en/latest/user_guide/basic_usage.html}{https://quadratik.readthedocs.io/en/latest/user\_guide/basic\_usage.html}. The value of $h$ has been fixed in advance in these examples. Recall that, in the general setting, we recommend that the optimal value of $h$ is obtained by the use of the function \texttt{select\_h}. See the simulation study in \cite{Markatou2023} for more details about the choice of the default setting here.  

\subsubsection{Normality Tests}

If $G$ denotes a distribution whose goodness of fit we wish to assess, the kernel-based quadratic distance can be employed for constructing a test statistic by measuring the distance between the sample data and the target distribution. This is known as the one-sample case. Let $G$ denote the $d$-dimensional normal distribution $N_d(\vect{0}, \mat{V})$. Consider $H_0: F=G$ versus $H_1:F\not=G$ and let $\vect{x}_1, \ldots, \vect{x}_n \sim F$ independent identically distributed (i.i.d.) observations, with cumulative distribution function (c.d.f.)  $\hat{F}$. 	By centering $K$ with respect to the distribution $G$, the KBQD can be written as \citep{Lindsay2014}
\begin{equation}
\label{eqn:center-lindsay2014}
    	d_K(\hat{F},G) = \iint K_{cen(G)}(\vect{s},\vect{t}) d\hat{F}(\vect{s}) d\hat{F}(\vect{t}). 
\end{equation}
Then, the corresponding $V$- and $U$-statistics are given by 
\begin{equation}
\label{eqn:u-stat}
    		V_{n} =  \frac{1}{n^2}\sum_{i=1}^n \sum_{j =i}^n K_{cen(G)}(\vect{x}_i,\vect{x}_j) \qquad\mbox{and}\qquad U_{n} =  \frac{1}{n(n-1)}\sum_{i=1}^n \sum_{j \not=i}^n K_{cen(G)}(\vect{x}_i,\vect{x}_j).
\end{equation}
Under the simple null hypothesis, the asymptotic distribution of $V_n$ is
\begin{equation*}
    nV_n \longrightarrow \sum_j \lambda_j Z_j^2  \qquad \mbox{ and } \qquad nU_n \longrightarrow \sum_j \lambda_j (Z_j^2 -1) \quad \mbox{ as } n \to \infty,
\end{equation*}
where $\lambda_j$s are the nonzero eigenvalues of the centered kernel $K_G$ under the null distribution G, and $Z_j$ are independent standard normal random variables. 
According to the Satterthwaite approximation, the asymptotic distribution of the statistic $nV_n$ is given by 
$$nV_n \sim c \cdot \chi^2_{DOF} \mbox{ as    } n\to \infty,$$ 		
with
\begin{equation*}
    c=\frac{trace(K^2_{cen(G)})}{trace(K_{cen(G)})} \qquad \mbox{ and } \qquad
    DOF(K_{cen(G)})=\frac{[trace(K_{cen(G)})]^2}{trace(K^2_{cen(G)})} ,
\end{equation*}
where
\begin{equation*}
    trace(K^2_{cen(G)}) = \left( |\Sigma_h|^{-\frac{1}{2}} |\Sigma_h + 4V|^{-\frac{1}{2}} - 2 |\Sigma_h + V|^{-\frac{1}{2}}|\Sigma_h+3V|^{-\frac{1}{2}} + |\Sigma_h + 2V|^{-\frac{1}{2}}\right),
\end{equation*}
and 
\begin{equation*}
    trace(K_{cen(G)}) = \left( |\Sigma_h|^{-\frac{1}{2}} -  |\Sigma_h + 2V|^{-\frac{1}{2}}\right).
\end{equation*}
Then, the cutoff value of the $nV_n$ statistic is determined by multiplying $c$ with the $95^{th}$ quantile of the chi-squared distribution $\chi^2_{\mathrm{DOF}}$.
To be able to obtain the critical value of the U-statistic is not necessary to know the eigen-decomposition of the centered kernel. We consider the standardized U-statistic $T_n = \frac{U_n}{\mathrm{Var}_G(U_n)}$, where, if $G=N_d(\vect{0},\mat{V})$,
\begin{equation*}
    \mathrm{Var}_G(U_n) = (2\pi)^{-d}\left( |\Sigma_h|^{-\frac{1}{2}} |\Sigma_h + 4V|^{-\frac{1}{2}} - 2 |\Sigma_h + V|^{-\frac{1}{2}}|\Sigma_h+3V|^{-\frac{1}{2}} + |\Sigma_h + 2V|^{-\frac{1}{2}}\right).
\end{equation*}
For this statistic, we consider the non-parametric calculation of the critical value presented in Algorithm \ref{alg:CV}, when $F_{\tau}=G$.\\
\begin{algorithm}[H]
\label{alg:CV}
 \caption{Computation of Critical Value}
\DontPrintSemicolon
Let $F_{\tau}$ denote the true (common) distribution under the null hypothesis. Let $B$ denote the number of sampling repetitions.   \\
\nlset{1} Generate i.i.d. observations $\vect{z}_1, \ldots,\vect{z}_{n_B}$ from the true distribution $F_{\tau}$ under $H_0$;\\
\nlset{2} Compute the KBQD test statistic;\\
\nlset{3} Repeat lines~\texttt{1}-\texttt{2} $B$ times; \\
\nlset{4} Order the obtained $B$ statistics from smallest to largest and select the statistic corresponding to the percentile given in the argument \texttt{Quantile}.
\end{algorithm}

Finally, notice that, for testing normality, if we want to assess $F=G$ with $G \sim N_d(\vect{\mu}, \mat{V})$ for some mean vector $\vect{\mu}$ and covariance matrix $ \mat{V}$, the normal kernel centered with respect to $G$ can be computed as 
\begin{align*}
K_{cen(G)}(\vect{s}, \vect{t}) = & K_{\mat{\Sigma_h}}(\vect{s}, \vect{t}) - K_{\mat{\Sigma_h} + \mat{V}}(\vect{\mu}, \vect{t}) \\
& -  K_{\mat{\Sigma_h} + \mat{V}}(\vect{s}, \vect{\mu}) + K_{\mat{\Sigma_h} + 2\mat{V}}(\vect{\mu}, \vect{\mu}). 
\end{align*}

\begin{example}[\textbf{Test for normality}]
\label{example:normality}

Here, we illustrate the usage of the introduced function for the normality test. We generate one sample from a multivariate standard normal distribution, that is $\vect{x} = (\vect{x}_1,\ldots,\vect{x}_n) \sim N_d(\vect{0},\mat{I}_d)$ with dimension $d=4$, and sample size $n=500$. Listing \ref{code:normality-R} shows the usage in \text{R}.   
\begin{lstlisting}[language=R, frame=tb, caption=R Code for the Normality test., label=code:normality-R]
> library(QuadratiK)
> library(mvtnorm)
> n <- 500
> d <- 4
> set.seed(2468)
> dat_norm <- rmvnorm(n, sigma = diag(d))

> h = 0.4
> set.seed(2468)
> system.time(norm_test <- kb.test(x=dat_norm, h=h, centering="Param"))
   user  system elapsed
   0.49    0.05    0.61 
> norm_test

 Kernel-based quadratic distance Normality test 
			 U-statistic 		 V-statistic 
--------------------------------------------
H0 is rejected: 	 FALSE 		 FALSE 
Test Statistic: 	 0.2595625 		 0.9979982 
Critical value (CV):	 1.601941 		 42.40783 
CV method:   
Selected tuning parameter h:  0.4 
\end{lstlisting}

The output of the test shows the provided value of $h$ as \texttt{Selected tuning parameter}. Recall that, if such a value is not provided, the algorithm for the selection of $h$ is automatically performed and the chosen value is displayed. 
Additionally, the package provides the \texttt{summary} method for the \texttt{kb.test} output object. Together with the results of the performed test, this function provides the qq-plots of each variable with a table of standard descriptive statistics. Listings \ref{code:normality_summary-R} shows the usage of the \texttt{summary} function, while the generated qq-plots are displayed in Figure \ref{fig:summary_norm}. 
\begin{lstlisting}[language=R, frame=tb, label=code:normality_summary-R, caption=R Code for the summary results of the normality test.]
> summary_norm <- summary(norm_test)

  Kernel-based quadratic distance Normality test 
    Test_Statistic Critical_Value Reject_H0
1        0.2595625      1.6601941     FALSE
2        0.9979982      42.407826     FALSE 
\end{lstlisting}
\begin{figure}[ht]
    \centering
    \includegraphics[width=0.68\linewidth]{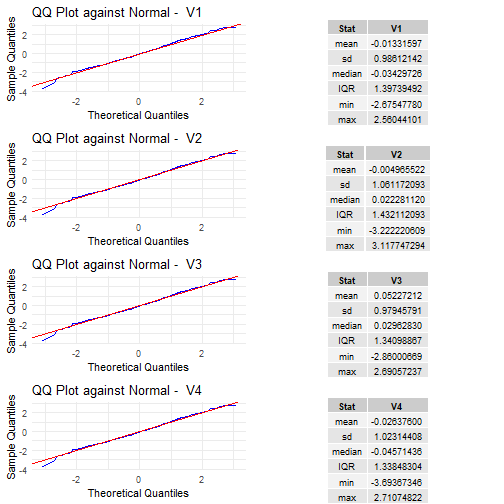}
    \caption{Figure automatically generated by the \texttt{summary} function on the result of the normality test. It displays the normal qq-plots (left) with a table of the standard descriptive statistics (right) for each variable.}
    \label{fig:summary_norm}
\end{figure}
\end{example}

\subsubsection[k-Sample Tests]{$k$-Sample Tests}
\label{subsec:k-sample}

The goodness-of-fit tests based on a kernel-based quadratic distance can be extended to the $k$-sample problem via the employment of a matrix distance, introduced by \cite{Markatou2023}. 

Consider the null hypothesis $H_0: F_1 = F_2 = \ldots = F_k$ against the alternative $H_1: F_i \not = F_j$, for some $1 \le i \not = j \le k$. Assume that under the null hypothesis $F_1 = \ldots = F_k = \bar{F}$, then the $ij$-th element of the matrix distance $D = (D_{ij})$ is defined as
\begin{equation*}
	D_{ij} = \iint K(\vect{x},\vect{y}) d(F_i - \bar{F})(\vect{x})d(F_j - \bar{F})(\vect{y}).
\end{equation*}
  Consider random samples of i.i.d. observations $\vect{x}^{(i)}_{1},\vect{x}^{(i)}_{2},\ldots, \vect{x}^{(i)}_{n_i} \sim F_i$, let $\hat{F}_i$ be the corresponding c.d.f. and $n = \sum_{i=1}^K n_i$. By centering the kernel function with respect to a chosen distribution $\bar{F}$, we have that
\begin{equation*}
	\hat{D}_{ij} = \iint K_{cen(\bar{F})}(\vect{x},\vect{y}) d\hat{F}_i(\vect{x})d\hat{F}_j(\vect{y}).
\end{equation*}
Then, the empirical version of the matrix distance $\hat{D}_n$ has as elements the $U$-statistics  
\begin{equation}
\label{eqn:mat-element-off}
	\hat{D}_{ij} = \frac{1}{n_i n_j} \sum_{l=1}^{n_i}\sum_{r=1}^{n_j}K_{cen(\bar{F})}(\vect{x}^{(i)}_l,\vect{x}^{(j)}_r) \qquad \mbox{ for }i \not= j
\end{equation}
and
\begin{equation}
\label{eqn:mat-element-diag}
	\hat{D}_{ii} = \frac{1}{n_i (n_i -1)} \sum_{l=1}^{n_i}\sum_{r\not= l}^{n_i}K_{cen(\bar{F})}(\vect{x}^{(i)}_l,\vect{x}^{(i)}_r) \qquad \mbox{ for }i = j.
\end{equation}

If we want to test the equality of $k$ samples against the general alternative $H_1: F_i \not = F_j$, for some $1 \le i \not = j \le k$, it is possible to construct an \textit{omnibus} test considering all the possible equalities included in the $k$-sample null hypothesis. This can be derived from the matrix distance as
\begin{equation}
\label{eqn:k-sample-stat}
	T_{n} = (K-1)\mathrm{trace}(\hat{D}_n) - 2 \sum_{i=1}^{K}\sum_{j> i}^{K}\hat{D}_{ij}.
\end{equation}
In our implementation, we provide the values of  $T_n$ and $\mathrm{Tr}_n = \mathrm{trace}(\hat{D}_n)$.

To be able to obtain the critical value of the considered statistics, the package provides the non-parametric calculation of the critical value, following Algorithm \ref{alg:CV}, when $F_{\tau}= \bar{F}$. Considering the pooled sample $\vect{z}_1, \ldots,\vect{z}_{n}$, the samples in step \texttt{1} of Algorithm \ref{alg:CV} are generated using one of the subsampling, bootstrap and permutation sampling procedures. 
Bootstrap and subsampling are two distinct methods for sampling data. Bootstrap involves drawing data points from the original dataset with replacement, creating new samples that are of the same size as the original dataset. On the other hand, subsampling creates new samples without replacement and typically these samples have smaller sample size, that is $n_B = b * n$ and $b \in (0,1]$. There is no clear guidance for the choice of the "optimal" subsample size $n_B$ and, the literature investigates this aspect according to optimal subsampling probabilities formulated by minimizing some function of the asymptotic distribution. For a more detailed discussion see \cite{politis2012subsampling} and \cite{yao_wang_2021}. Finally, the permutation algorithm generates new samples of the same sample size drawing from the observations in the original data set. Notice that this procedure coincides with the subsampling algorithm with $b=1$, where new samples with the same sample size of the original data set are generated by sampling without replacement.
\cite{Markatou2023} investigated the algorithm for computing the critical value with respect to the chosen sampling algorithm and the number of repetitions $B$. The permutation and bootstrap sampling methods showed higher and more stable performance, in terms of level and power, for high dimensions and for low sample size. The subsampling algorithm is preferred in case of large sample size, since generating samples with smaller sample sizes requires significantly less computational time. It is worth noting that the choice of the number of bootstrap/subsampling/permutation replications $B$ can have an evident influence in terms of performance.  Previous simulation studies indicate that varying the value of $B$ does not consistently impact the kernel-based quadratic distance tests, in terms of both level and power. While smaller values of $B$ may achieve slightly higher power, they can also introduce instability in the results, potentially leading to different conclusions in the goodness-of-fit problem. A larger $B$ can significantly increase the computational time. In our implementation, we suggest $B=150$ since it provides a balance between satisfactory statistical power while maintaining result stability and practical computational time.  However, note that users are given the flexibility to adjust the value of $B$ based on the specific needs and computational resources.   

For the $k$-sample problem, we would prefer not to make any distributional assumption on the (unknown) common distribution $\bar{F}$, that is $F_1 = \ldots,F_K = \bar{F}$. In order to construct non-parametric KBQD test statistics, the kernel function is centered with respect to the weighted average distribution $\bar{F} = \frac{1}{n} \sum_{i=1}^K n_i F_i$, with $n=\sum_{i=1}^K n_i$. Let $\mat{z}$ denote the pooled sample, then the non-parametric centered kernel can be computed as 
\begin{align*}
    K_{cen(\bar{F})}(\vect{x},\vect{y}) = &
    K(\vect{x},\vect{y}) - \frac{1}{n}\sum_{i=1}^{n} K(\vect{x},\vect{z}_i) - \frac{1}{n}\sum_{i=1}^{n} K(\vect{z}_i,\vect{y}) \\
    & + \frac{1}{n(n-1)}\sum_{i=1}^{n} \sum_{j \not=i}^{n} K(\vect{z}_i,\vect{z}_j).
\end{align*}

\begin{example}[\textbf{$k$-sample test}]

We generated three samples, with $n=200$ observations each, from a 2-dimensional Gaussian distributions with mean vectors $\vect{\mu}_1 = (0, \frac{\sqrt{3}}{3})$, $\vect{\mu}_2 = (-\frac{1}{2}, -\frac{\sqrt{3}}{6})$ and $\vect{\mu}_2 = (\frac{1}{2}, -\frac{\sqrt{3}}{6})$, and the Identity matrix as covariance matrix. In this situation, the generated samples are well separated, following different Gaussian distributions, i.e. $\vect{X}_1 \sim N_2(\vect{\mu}_1, \mat{I})$, $\vect{X}_2 \sim N_2(\vect{\mu}_2, \mat{I})$ and $\vect{X}_3 \sim N_2(\vect{\mu}_3, \mat{I})$. Listings \ref{code:ksample-R} illustrates the usage, in \texttt{R}, of the \texttt{kb.test} function for the $k$-sample problem. It requires the vector \texttt{y} indicating the membership to groups. 
\begin{lstlisting}[language=R, caption=R Code for the k-sample test., label=code:ksample-R, frame=tb]
> sizes <- rep(200,3)
> eps = 1
> set.seed(2468)
> x1 <- rmvnorm(sizes[1], mean = c(0,sqrt(3)*eps/3))
> x2 <- rmvnorm(sizes[2], mean = c(-eps/2,-sqrt(3)*eps/6))
> x3 <- rmvnorm(sizes[3], mean = c(eps/2,-sqrt(3)*eps/6))
> x <- rbind(x1, x2, x3)
> y <- as.factor(rep(c(1,2,3), times=sizes))

> h=1.5
> set.seed(2468)
> system.time(k_test <- kb.test(x=dat_k, y=y_k, h=h))
   user  system elapsed 
   1.27   0.19    1.47 
> k_test

 Kernel-based quadratic distance k-sample test 
U-statistics	 Dn 		 Trace 
------------------------------------------------
Test Statistic:	 11.844 	 38.6817 
Critical Value:	 0.5623288 	 1.836868 
H0 is rejected:	 TRUE 		 TRUE 
CV method:  subsampling 
Selected tuning parameter h:  1.5 
\end{lstlisting}
Recall that the computed test statistics correspond to $T_n$ and $\mathrm{Tr}_n$, given in equation (\ref{eqn:k-sample-stat}). 
When the $k$-sample test is performed, the \texttt{summary} method on the \texttt{kb.test} object returns the results of the tests together with the standard descriptive statistics for each variable computed, overall and with respect to the provided groups. The usage of the \texttt{summary} function is shown in listings \ref{code:ksample-summary}.
\begin{lstlisting}[language=R, frame=tb, label=code:ksample-summary, caption=R Code for the summary results of the k-sample test.]
> summary_ktest <- summary(k_test)

 Kernel-based quadratic distance k-sample test 
  Statistic Test_Statistic Critical_Value Reject_H0
1        Dn        11.8440      0.5623288      TRUE
2     Trace        38.6817      1.8368685      TRUE
> summary_ktest$summary_tables
[[1]]
            Group 1    Group 2    Group 3       Overall
mean   -0.005959147 -0.5370127  0.5442058  0.0004113282
sd      0.997319811  0.9583059  1.0374834  1.0900980006
median -0.028244038 -0.5477108  0.5297478 -0.0239486027
IQR     1.478884929  1.4105832  1.4234532  1.5377418198
min    -2.860006689 -3.1869808 -2.2119189 -3.1869807848
max     2.151784802  2.0647648  3.1580700  3.1580700259

[[2]]
          Group 1    Group 2    Group 3     Overall
mean    0.4935364 -0.4042219 -0.2461729 -0.05228613
sd      1.0449582  1.0411639  1.0474989  1.11391575
median  0.5281635 -0.4325995 -0.2950922 -0.09520111
IQR     1.4001089  1.4662111  1.2867345  1.48444495
min    -2.6448703 -2.8786352 -3.4932849 -3.49328492
max     3.0792766  2.6788424  2.8290722  3.07927659
\end{lstlisting}
\end{example}

\subsubsection{Two-sample Tests}

For the special case $K=2$, the matrix distance gives the test statistic
\begin{align*}
    D_{{n_1},n_2} 
=  &\frac{1}{{n_1}({n_1}-1)}\sum_{i=1}^{n_1} \sum_{j \not=i}^{n_1} K_{cen(\bar{F})}(\vect{x}_i,\vect{x}_j) - \frac{2}{{n_1}{n_2}}\sum_{i=1}^{n_1} \sum_{j =1}^{n_2} K_{cen(\bar{F})}(\vect{x}_i,\vect{y}_j) \\
    & + \frac{1}{{n_2}({n_2}-1)}\sum_{i=1}^{n_2} \sum_{j \not=i}^{n_2} K_{cen(\bar{F})}(\vect{y}_i,\vect{y}_j)
\end{align*}
and the corresponding trace statistic. For further theoretical information on the two- and $k$-sample tests see \cite{Markatou2023}.

The two-sample test can be additionally performed by providing the two sample to be compared as \texttt{x} and \texttt{y}. This option is showed in the next example.

\begin{example}[\textbf{Non-parametric two-sample test}]

We generate the sample $\vect{y} = (\vect{y}_1, \ldots,\vect{y}_n)$ from a skew-normal distribution $SN_d(\vect{0},\mat{I}_d, \vect{\lambda})$, where $d=4$, $n=200$ and $\vect{\lambda}= \lambda(1,\ldots,1)$ with $\lambda = 0.5$. Listing \ref{code:two-sample-R} shows the codes using the \texttt{kb.test} for testing if $\vect{y}$ and $\vect{x}$, from Example \ref{example:normality}, follow the same distribution. 
\begin{lstlisting}[language=R, frame=tb, label=code:two-sample-R, caption=R Code for the two-sample test.]
> library(sn)
> n <- 200
> d <- 4
> skewness_y <- 0.5
> set.seed(2468)
> x_2 <- rmvnorm(n, mean = rep(0,d))
> y_2 <- rmsn(n=n, xi=0, Omega = diag(d), alpha=rep(skewness_y,d))
> h = 2
> set.seed(2468)
> system.time(two_test <- kb.test(x=x_2, y=y_2, h=h))
   user  system elapsed 
   0.43    0.00    0.45 
> two_test

 Kernel-based quadratic distance two-sample test 
U-statistics	 Dn 		 Trace 
------------------------------------------------
Test Statistic:	 4.276823 	 9.843008 
Critical Value:	 0.7745576 	 1.783123 
H0 is rejected:	 TRUE 		 TRUE 
CV method:  subsampling 
Selected tuning parameter h:  2
\end{lstlisting}
For the two-sample case, the \texttt{summary} function provides the results from the test and the descriptive statistics per variable and per group, as similarly described for the $k$-sample test. Additionally, it generates the qq-plots comparing the quantiles of the two groups for each variable. 
Listing \ref{code:2sample-summary} shows the \texttt{summary} function, while Figure \ref{fig:summary_two} shows the generated qq-plots for the example data set.
\begin{lstlisting}[language=R, frame=tb, label=code:2sample-summary, caption=R Code for the summary results of the two-sample test.]
> summary_two <- summary(two_test)

 Kernel-based quadratic distance two-sample test 
  Statistic Test_Statistic Critical_Value Reject_H0
1        Dn       4.276823      0.7745576      TRUE
2     Trace       9.843008      1.7831227      TRUE
> summary_two$summary_tables
[[1]]
            Group 1    Group 2    Overall
mean    0.021762263  0.3799990  0.2008806
sd      1.014655344  0.9498167  0.9977884
median -0.005110155  0.3833061  0.2125618
IQR     1.471877262  1.1310211  1.3666010
min    -2.675477796 -2.2219439 -2.6754778
max     2.300153117  3.1690406  3.1690406

[[2]]
           Group 1    Group 2     Overall
mean   -0.03347117  0.2216529  0.09409085
sd      1.06408749  1.0304067  1.05383755
median  0.02476594  0.1717768  0.09272994
IQR     1.52458343  1.3739349  1.45668193
min    -3.22222061 -2.6162342 -3.22222061
max     2.96751758  2.3300745  2.96751758

[[3]]
           Group 1    Group 2    Overall
mean   -0.06473408  0.3312699  0.1332679
sd      0.93818786  0.9868499  0.9818422
median -0.07044427  0.4006745  0.1382735
IQR     1.37135831  1.2185714  1.3854150
min    -2.86000669 -3.0246026 -3.0246026
max     2.56476485  2.7590501  2.7590501

[[4]]
          Group 1    Group 2     Overall
mean   -0.1658894  0.2065222  0.02031639
sd      1.0175325  0.9718613  1.01104987
median -0.2371959  0.1427746  0.04889195
IQR     1.3802070  1.2320445  1.32957715
min    -2.5899601 -2.0159679 -2.58996007
max     2.7066430  2.6637589  2.70664302
\end{lstlisting}
\begin{figure}[htbp]
    \centering
    \includegraphics[width=0.78\linewidth]{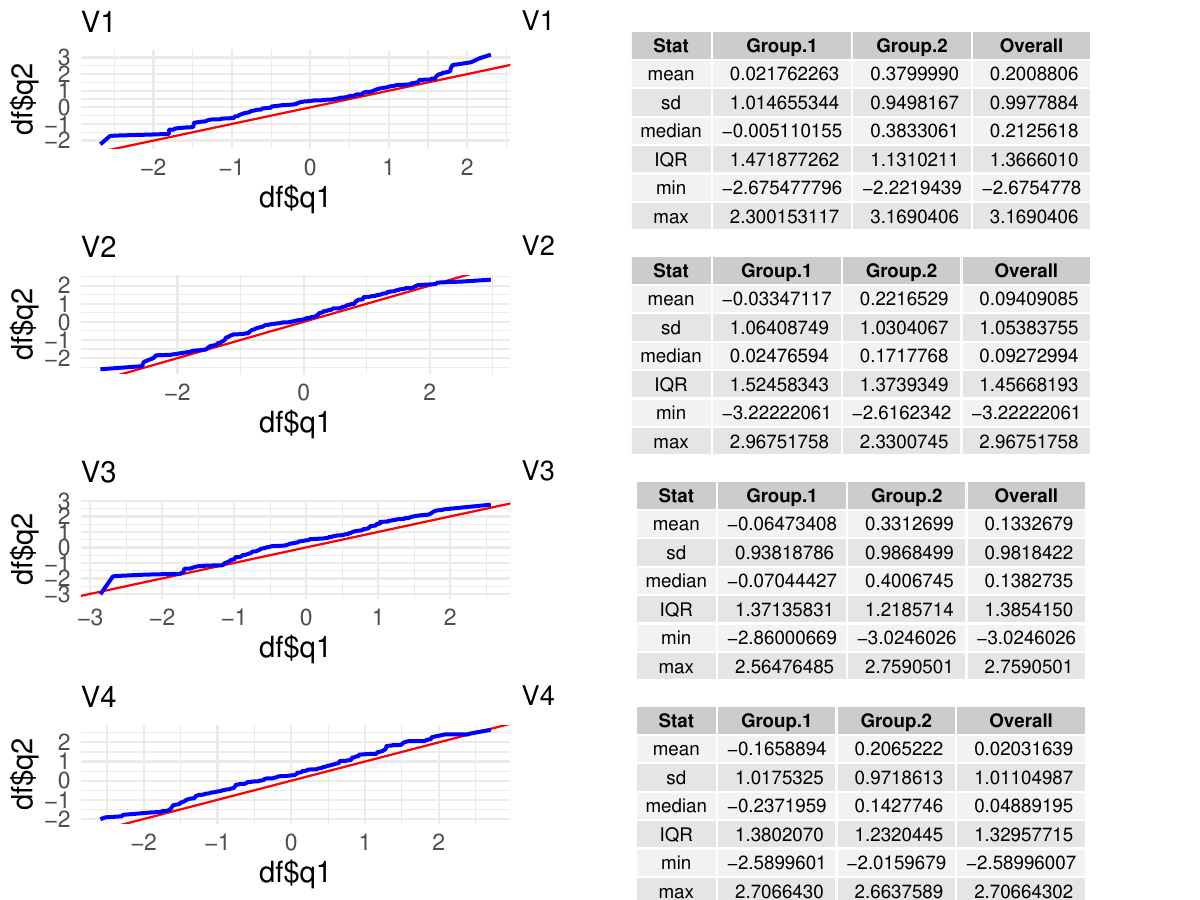}
    \caption{Figure automatically generated by the \texttt{summary} function on the result of the two-sample test. It displays the  qq-plots between the two samples (left) with a table of the standard descriptive statistics for each variable (right), computed per group and overall.}
    \label{fig:summary_two}
\end{figure}
\end{example}


\subsection{Selection of the tuning parameter}
\label{subsec:select-h}

In this section, we present the algorithm for selecting the tuning parameter \texttt{h} of the normal kernel. This procedure, proposed by \cite{Markatou2023}, properly adapts the strategy based on the mid-power analysis introduced by \cite{Lindsay2014} to the $k$-sample problem.

The power of the test statistics can be approximated using asymptotic normality under the alternative through the power function   
\begin{equation*}
    \beta(H_1) = \mathbb{P}(T_{n}> z_\alpha \sigma_0) = \mathbb{P}\left(Z> \frac{z_\alpha \sigma_0 - \mathbb{E}_{H_1} (T_{n})}{\sigma_1}\right), 
\end{equation*}
where $z_\alpha$ is the $\alpha$-quantile of the normal distribution with level $\alpha$, $\sigma_0^2$ and $\sigma_1^2$ are the variances of the test statistic $T_{n}$ under the null and the alternative hypotheses, respectively. We focus on the mid-power manifold $\beta(H_1)=0.5$, which allows for a simpler analytic calculation. The algorithm used to select the tuning parameter $h$ is described in the following steps. 
\begin{itemize}
    \item Identify the family of target alternatives $\{F_\delta\}$, $\delta \ge 0$. The case $\delta = 0$ corresponds to $H_0$, while larger values of $\delta$ indicate more evident departures from the null hypothesis.
    \item For each value of $h$, define the smallest value of $\delta$ for which the test achieves power $0.5$ as
    \begin{equation*}
        \delta_{mid}(h) = \arg\min\{\delta:\beta_h(\delta)=0.5\}.
    \end{equation*}
    \item The \textit{optimal} tuning parameter $h^\ast$ is defined as the value of $h$ with the smallest mid-power sensitivity, that is
    \begin{equation*}
        h^\ast = \arg\min_h\{\delta_{mid}(h)\}.
    \end{equation*}
\end{itemize}
The implementation of the described steps is reported in Algorithm \ref{alg:select-h}.\\ 
A fundamental point is the specification of an alternative family of distributions. Let this be defined by $F_\delta(\mu, \Sigma, \lambda)$ where $\mu, \Sigma, \lambda$ indicate the location, covariance and skewness parameters of the distribution. Examples of targeted alternatives are provided after the description of Algorithm \ref{alg:select-h}.

\begin{algorithm}[H]
\caption{Selection of tuning parameter}
\label{alg:select-h}
\DontPrintSemicolon
\nlset{1}Compute the estimates of mean $\hat{\mu}$, covariance matrix $\hat{\Sigma}$ and skewness $\hat{\lambda}$ from the pooled sample.  \\
\nlset{2}Choose the family of alternatives $F_\delta = F_\delta(\hat{\mu},\hat{\Sigma}, \hat{\lambda})$. \\
\For{$\delta$}{
    \For{$h$}{
	\nlset{3}Generate $\vect{X}_1,\ldots,\vect{X}_{K-1}  \sim F_0$, for $\delta=0$;\\
 	\nlset{4}Generate $\vect{X}_K \sim F_\delta$;\\
	\nlset{5}Compute the $k$-sample test statistic between $\vect{X}_1, \vect{X}_2, \ldots, \vect{X}_K$ with kernel parameter $h$;\\
	\nlset{6}Repeat lines~3-5 $N$ times. \\
	\nlset{7}Compute the power of the test. If it is greater than 0.5, select $h$ as optimal value. 
	}
}
\nlset{8}If an optimal values has not been selected, choose the $h$ which corresponds to maximum power. 
\end{algorithm}
Step $\texttt{2}$ of the algorithm requires that the family of alternatives $\{F_\delta\}$ must be identified. In order to properly handle the data in hand, the proposed implementation of the algorithm includes the following target alternatives, with corresponding default values of $\delta$:
\begin{itemize}
\item[(i)] \textit{location} alternatives $F_\delta = SN_d(\hat{\mu} + \delta,\hat{\Sigma}, \hat{\lambda})$,with $\delta = 0.2, 0.3, 0.4$; 
\item[(ii)] \textit{scale} alternatives $F_\delta = SN_d(\hat{\mu} ,\hat{\Sigma}*\delta, \hat{\lambda})$, $\delta = 0.1, 0.3, 0.5$; 
\item[(iii)] \textit{skewness} alternatives, $F_\delta = SN_d(\hat{\mu} ,\hat{\Sigma}, \hat{\lambda} + \delta)$, with $\delta = 0.2, 0.3, 0.6$.
\end{itemize}
The following values of $h = 0.6, 1, 1.4, 1.8, 2.2$ and $N=50$ are set as default. The algorithm is implemented through the function \texttt{select\_h}. The values of $\delta$ and $h$ can be set by the user, if a more extensive search is required. \cite{Markatou2023} conducted an extensive simulation study considering location and skewness alternatives for the two-sample and $k$-sample problems. Default values here are chosen considering the results from this simulation study.
In case of data following different distributions, we suggest a more detailed grid search providing more values for $h$, and $\delta$. This is the recommended practice whenever the computational resources are available. Also, the \texttt{select\_h} function allows users to specify the number of cores to be used for parallelization using the \texttt{n\_cores} argument. In resource-scarce environments, the \texttt{n\_cores} can be set to low values, including 1, which means that the function will execute sequentially. Meanwhile, in environments where multiple cores are available, users can increase \texttt{n\_cores} to a values greater than the default of 2 cores to take advantage of increased parallelization. 
The function \texttt{select\_h} takes as inputs the data matrix \texttt{x}, vector of labels \texttt{y}, and the type of alternatives (one of \textit{"location"}, \textit{"scale"} or \textit{"skewness"}). Similarly to the function \texttt{kb.test}, the algorithm for the selection of $h$ for the two-sample test can be performed providing the two samples as \texttt{x} and \texttt{y}. \texttt{select\_h} returns not only the selected value of $h$, but also the power plot versus the considered list of $h$ values for each tested value of $\delta$. This provides a set of possible values of $h$ with high power performance. The following examples show the usage of these functions considering the data sets previously generated for the two-sample and k-sample problems. 

\begin{example}[$k$-sample test -- Continued]
We consider the three 2-dimensional samples generated in Example 3.2, with $n=200$ observations each. Listing \ref{code:select_h_k-R} shows the codes for the usage of the function \texttt{select\_h} for the $k$-sample tests. This function needs the input \texttt{x} and \texttt{y} as the function \texttt{kb.test} for the $k$-sample problem. Figure \ref{fig:power_h_k} shows the generated power plot.     
\begin{lstlisting}[language=R, frame=tb, label=code:select_h_k-R, caption=R Code for the the selection of $h$.]
> y <- as.numeric(y_k)
> set.seed(2468)
> time_h_k <- system.time( h_k <- select_h(x=dat_k, y=y, 
                        alternative="skewness"))
> time_h_k
   user  system elapsed 
   0.30   0.00   323.42 
> h_k$h_sel
[1] 0.8
\end{lstlisting}
\begin{figure}[htb]
\centering
    \includegraphics[width=0.75\linewidth]{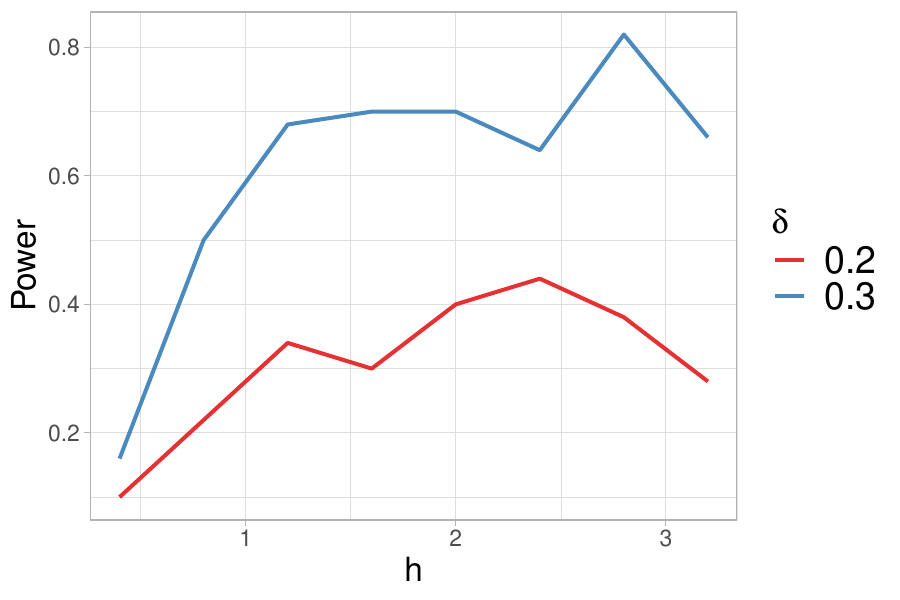}
    \caption{Plot generated by the \texttt{select\_h} function on the result of the selection of $h$ algorithm for the $k=3$ two-dimensional samples, with size $n=200$, in Example 3.2. It displays the obtained power versus the considered $h$, for each value of skewness alternative $\delta$ considered.}
    \label{fig:power_h_k}
\end{figure}
As it is seen from Figure \ref{fig:power_h_k}, when the alternative distribution $F_\delta$ with $\delta=0.2$ is considered, there are no values of $h$ within the indicated range which achieve power greater than or equal to 0.5. For the second value of $\delta=0.3$, $h=0.8$ is chosen as optimal value since it is the smallest value with power greater than 0.5. Additionally, it gives a possible set of values with high power performance.
\end{example}

\begin{example}[Two-sample test -- Continued]
We consider the two samples generated in Example 3.3, with sample size $n=200$ and dimension $d=4$. Listing \ref{code:select_h-R} shows the codes for the usage of the function \texttt{select\_h} for the two-sample tests. This function needs the input \texttt{x} and \texttt{y} as the function \texttt{kb.test} for the two-sample problem. Figure \ref{fig:power_h_2} shows the generated power plot.    
\begin{lstlisting}[language=R, frame=tb, label=code:select_h-R, caption=R Code  for the the selection of $h$ for the two-sample test.]
> set.seed(2468)
> time_h_2 <- system.time( h_test2 <- select_h(x=x_2, y=y_2, alternative="location"))
> time_h_2
   user  system elapsed
   0.14   0.00   54.92  
> h_test2$h_sel
[1] 1.2
\end{lstlisting}
\begin{figure}[htb]
    \centering
    \includegraphics[width=0.75\linewidth]{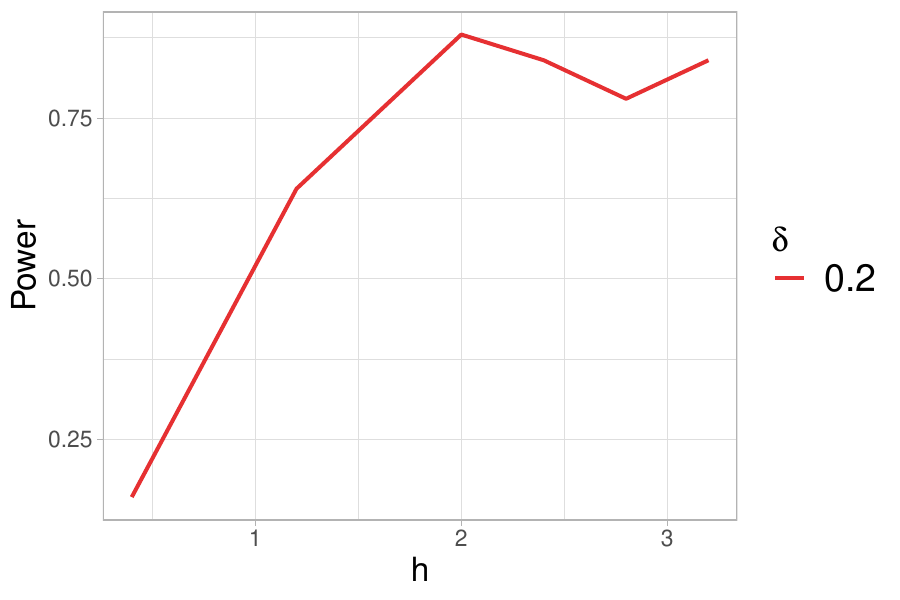}
    \caption{Plot generated by the \texttt{select\_h} function on the result of the selection of $h$ algorithm on the two-sample data set in Example 3.3, with sample size $n=200$ and dimension $d=4$. It displays the obtained power versus the considered $h$, for the alternative $\delta=0.2$.}
    \label{fig:power_h_2}
\end{figure}
For the two-sample data set, the power analysis with respect to the alternative distribution with $\delta=0.2$ returns a set of possible values for $h$ which achieve high power, as showed in Figure \ref{fig:power_h_2}. Then, different values of $\delta$ are not explored.

Usually, there are more than one unique value of $h$ that achieve power greater than or equal to 0.5. In this case, depending on the number of runs $N$, it is possible for the algorithm to identify one or more values of $h$ that are appropriate to use since they offer power greater than 0.5.  
\end{example}

\subsection{Real data examples}
\label{subsec:data-ksample}

In this section, we show the application of the provided two and $k$-sample kernel-based quadratic distance tests in two real data applications. 

\subsubsection{Breast Cancer Wisconsin Data}

We first consider the Breast Cancer Wisconsin Data \citep{breastCancer}, publicly available from the UCI Machine Learning Repository's website. The data includes features computed from digitized images of biopsies, of a fine needle aspirate (FNA) of a breast mass. Ten real-valued features are computed for each cell nucleus, describing characteristics present in the image (that is radius, texture, perimeter, area, smoothness, compactness (perimeter$^2$ / area - 1.0), concavity, concave points, symmetry, fractal dimension. The mean, standard error and “worst” or largest (mean of the three largest values) of these features
were computed for each image, resulting in $d=30$ variables. Over the total $n=569$ observations, 357 are labeled as benign (B) and the remaining 212 as malignant (M). Listing \ref{code:breast-cancer-data} shows the usage of the \texttt{kb.test} and \texttt{select\_h} functions for the breast cancer data set.  
\vspace{-3mm}
\begin{lstlisting}[language=R, frame=tb, label=code:breast-cancer-data, caption=R Code for the breast cancer Winsconsin data example.]
> dat <- read.csv("Wisconsin_Diagnostic.csv")
> x <- dat[which(dat$y=="B"),-ncol(dat)]
> y <- dat[which(dat$y=="M"),-ncol(dat)]
> # Normalize observations
> x <- x/sqrt(rowSums(x^2))
> y <- y/sqrt(rowSums(y^2))

> # Perform algorithm for selection of h
> time_wisc <- system.time( h_sel <- select_h(x = x, y = y, alternative = "skewness",
                                            method = "subsampling", b = 0.5))
> time_wisc
   user  system elapsed 
   0.22    0.04  138.20
> h_sel$h_sel
[1] 0.4
> # Perform the test
> system.time(kbqd_test <- kb.test(x = x, y = y, h = h_sel$h_sel))
   user  system elapsed
   1.28    0.34    1.66 
> kbqd_test

 Kernel-based quadratic distance two-sample test 
U-statistics	 Dn 		 Trace 
------------------------------------------------
Test Statistic:	 11.57605 	 103.1909 
Critical Value:	 0.09466646 	 0.8428861 
H0 is rejected:	 TRUE 		 TRUE 
CV method:  subsampling 
Selected tuning parameter h:  0.4
\end{lstlisting}

\subsubsection{Wine Data}

Next, we consider the Wine Data \citep{misc_wine_109}, available from the UCI Machine Learning Repository's website. These data are the results of a chemical analysis of wines grown in the same region in Italy but derived from three different cultivars. The analysis determined the quantities of 13 constituents (Alcohol, Malic acid, Ash, Alcalinity of ash, Magnesium, Total phenols, Flavanoids, Nonflavanoid phenols, Proanthocyanins, Color intensity, Hue, OD280/OD315 of diluted wines, Proline) found in each of the three types of wines. This data, usually considered in classification context, consists of 178 observations. Listing \ref{code:wine-data} reports the codes for performing the $k$-sample kernel based quadratic distance tests for comparing the samples coming from different cultivars in the Wine data, using the \texttt{kb.test} and \texttt{select\_h} functions. 
\begin{lstlisting}[language=R, frame=tb, label=code:wine-data, caption=R Code for the Wine data example.]
> wine <- read.csv("Wine.csv")
> # select the data and the labels 
> x <- wine[,-14]
> x <- x/sqrt(rowSums(x^2))
> y <- factor(wine[,14])
> # Perform the algorithm for the selection of h 
> time_wine <- system.time(h_sel_wine <- select_h(x = x, y = y, alternative = "skewness", method = "subsampling", b = 0.5))
> time_wine
   user  system elapsed 
   0.33    0.05   24.56 
> h_sel_wine$h_sel
[1] 1.6
> # Perform the k-sample test
> system.time(kbqd_test_wine <- kb.test(x = x, y = y, h = h_sel_wine$h_sel))
   user  system elapsed
   0.14    0.00    0.17 
> kbqd_test_wine

 Kernel-based quadratic distance k-sample test 
U-statistics	 Dn 		 Trace 
------------------------------------------------
Test Statistic:	 8.491507 	 37.88043 
Critical Value:	 0.3321461 	 1.485128 
H0 is rejected:	 TRUE 		 TRUE 
CV method:  subsampling 
Selected tuning parameter h:  1.6
\end{lstlisting}
 
\subsection{The function pk.test}
\label{subsec:test_uniform}

The package offers also a method for analysis of spherical data, i.e. $\vect{x}_1, \ldots,\vect{x}_n \in S^{d-1}$. In this case, the Poisson kernel appears to be the natural generalization of the normal distribution on the $d$-dimensional sphere.  For $\vect{u},\vect{v} \in {S}^{d-1}$ and concentration parameter $\rho$, with $0 < \rho< 1$, the Poisson kernel is given as 
\begin{equation}
\label{eqn:poisson-kernel}
K_{{\rho}}(\vect{u}, \vect{v}) = \frac{1-\rho^2}{(1+\rho^2 -2\rho (\vect{u} \cdot \vect{v}))^{\frac{d}{2}}}.
\end{equation}
The Poisson kernel is the basis for the development of test statistics that assess the goodness-of-fit of $H_0: F=G$, where $G=\mathcal{U}(S^{d-1})$ denotes the Uniform distribution on the sphere. It belongs to the class of diffusion kernels, and the Poisson kernel centered with respect to the uniform distribution $\mathcal{U}(S^{d-1})$ can be computed as 
\begin{equation}
\label{eqn:poisson-center}
K_{cen(G)}(\vect{u}, \vect{v}) = K_{\rho}(\vect{u}, \vect{v}) - 1.
\end{equation}

The function \texttt{pk.test} performs the kernel-based quadratic distance test for uniformity on the sphere using the Poisson kernel developed by \cite{Ding2023}. Using the centered Poisson kernel given in equation (\ref{eqn:poisson-center}), the following two test-statistics are computed:
\begin{equation*}
    T_{n}=\frac{U_{n}}{\sqrt{Var(U_{n})}}, \mbox{ with } 
Var(U_{n})= \frac{2}{n(n-1)}\left[\frac{1+\rho^{2}}{(1 -\rho^{2})^{d-1}}-1\right],
\end{equation*}
and
\begin{equation*}
    S_{n}=\frac{1}{n}\sum_{i=1}^{n}\sum_{j=1}^{n}K_{cen(G)}(\mathbf{x}_{i}, \mathbf{x}_{j}).
\end{equation*}
The asymptotic distribution of the statistic $S_n$ is given by 
$$S_n \sim c \cdot \chi^2_{DOF} \mbox{ as    } n\to \infty,$$ 		
where
\begin{equation*}
    c=\frac{trace(K^2_{cen(G)})}{trace(K_{cen(G)})} = \frac{(1+\rho^{2})-(1-\rho^{2})^{d-1}}{(1+\rho)^{d}-(1-\rho^{2})^{d-1}}
\end{equation*}
and
\begin{equation}
\label{eqn:DOF}
    DOF(K_{cen(G)})=\frac{[trace(K_{cen(G)})]^2}{trace(K^2_{cen(G)})} =\left(\frac{1+\rho}{1-\rho} \right)^{d-1}\left\{ \frac{\left(1+\rho-(1-\rho)^{d-1} \right )^{2}}{1+\rho^{2}-(1-\rho^{2})^{d-1}}\right \}.
\end{equation}
Then, the cutoff value of the $S_n$ statistic is determined by multiplying $c$ with the $95^{th}$ quantile of the chi-squared distribution $\chi^2_{\mathrm{DOF}}$.  The critical value of statistic $T_n$ is obtained empirically, following Algorithm \ref{alg:CV}, with $F_\tau = \mathcal{U}(S^{d-1})$. The function takes the following arguments.
\begin{itemize}
    \item \texttt{x}: numeric matrix of data values.
    \item \texttt{rho}: concentration parameter.
    \item \texttt{B}: the number of iterations to use for critical value estimation, B = 300 as default.
    \item \texttt{Quantile}: the quantile to use for critical value estimation, 0.95 is the default value.
\end{itemize}

\begin{example}[\textbf{Uniformity test on the Sphere}]
We generated $n=200$ observations from the uniform distribution on $S^{d-1}$, with $d=3$, and $\rho = 0.7$. Listing \ref{code:uniform-R} illustrates the usage of the \texttt{pk.test} function for testing uniformity of the generated sample. 

\begin{lstlisting}[language=R, frame=tb, label=code:uniform-R, caption=R Code for the Uniformity test on the Sphere.]
> library(QuadratiK)
n <- 200
d <- 3
set.seed(2468)
z <- matrix(rnorm(n * d), n, d)
dat_sphere <- z/sqrt(rowSums(z^2))
> rho = 0.7
> set.seed(2468)
> system.time(res_unif <- pk.test(x=dat_sphere, rho=rho))
   user  system elapsed 
   0.96    0.00    1.08 
> res_unif

 Poisson Kernel-based quadratic distance test of Uniformity on the Sphere 
Selected consentration parameter rho:  0.7 

U-statistic:

H0 is rejected:  FALSE 
Statistic Un:  -0.9756673 
Critical value:  1.725683 

V-statistic:

H0 is rejected:  FALSE 
Statistic Vn:  14.89598 
Critical value:  23.22949 
\end{lstlisting}
As for the test for normality, the \texttt{summary} method for the \texttt{pk.test} output object provides the results of the performed test, and generates a figure showing the qq-plots against the uniform distribution of each variable with a table of standard descriptive statistics. Listing \ref{code:uniform_summary-R} shows the usage of the \texttt{summary} function, while the generated qq-plots are not displayed here. 
\begin{lstlisting}[language=R, frame=tb, label=code:uniform_summary-R, caption=R Code for the Uniformity test on the Sphere.]
> summary_unif <- summary(res_unif)

 Poisson Kernel-based quadratic distance test of Uniformity on the Sphere 
  Test_Statistics Critical_Value Reject_H0
1      -0.9756673     1.725683       FALSE
2      14.8959834    23.22948694     FALSE
\end{lstlisting}
\end{example}

\section{Clustering on the Sphere} 
\label{sec:clustering}

In this section, we consider the problem of clustering data that reside on the $d$-dimensional sphere. Examples of data that live on such spaces are directional data and data that can be transformed into directions. Computational methods for clustering directional data have been proposed in the literature. Non-parametric approaches to clustering include algorithms such as K-means clustering \citep{Ramler2008, Maitra2010}, spherical K-means \citep{Dhillon2001} and online spherical K-means \citep{Zhong2005}. In particular, spherical K-means corresponds to a version of K-means using the cosine similarity in place of the Euclidean distance. 

\cite{Banerjee2005} proposed a clustering method based on a finite mixture of von Mises--Fisher (vMF) distributions for clustering genomic and text data. 
Note that, the spherical K-means algorithm has been shown to be a special case of a generative model based on a mixture of vMF distributions with equal concentration parameters and priors for the components \citep{Banerjee2002}. 

While most of the clustering methods in the literature perform well on small and medium dimensions data sets, high-dimensional data are less frequently addressed. A promising alternative for dealing with high-dimensional data is kernel-based methods. In this context, \cite{Ng2001} introduced the spectral clustering, which relies on the top eigenvectors of an affinity matrix. 
Furthermore, \cite{KIM2020} recently proposed improvements on the spherical $k$-means to overcome the issues related to dimensionality growth in the context of document clustering.
 
As shown in section \ref{subsec:test_uniform}, the Poisson kernel density offers a natural way of assessing goodness of fit for spherical (or spherically transformed) data. The $d$-dimensional Poisson kernel in equation (\ref{eqn:poisson-kernel}) is a density function with respect to the uniform measure on the sphere, written as 
\begin{equation}
\label{eqn:poisson-d-density}
f(\vect{x}| \rho, \vect{\mu}) = \frac{1 - \rho^2}{\omega_d(1 + \rho^2 - 2 \rho \vect{x}\vect{\mu})^{d/2}},
\end{equation}
where $\omega_d = 2\pi^{d/2}\{\Gamma(\frac{d}{2})\}^{-1}$ is the surface area of the unit sphere, $||\vect{\mu}|| =1$ is the vector orienting the center of the distribution and the concentration parameter $ 0 < \rho < 1$ is related to the variance of the distribution. When $\rho \to 0$, the Poisson kernel-based density tends to the uniform density on the sphere. Recall that the parameter $\rho$ is directly connected to the DOF of the corresponding kernel-based test given in equation (\ref{eqn:DOF}) \citep{Ding2023}. In the univariate case, the Poisson-kernel based density is 
the circular density also known as wrapped Cauchy density, while in dimension $d$ it is obtained as the
density of the “exit” distribution on the sphere \citep{kato2013}.

\subsection{Clustering Algorithm}

The clustering algorithm, presented as Algorithm \ref{alg:clustering} below, proposed by \cite{golzy2020} consists of a parametric mixture model approach based on Poisson kernel-based distributions on the unit sphere. 

\begin{algorithm}[htbp]
\caption{Poisson Kernel-Based Clustering on the Sphere}
\label{alg:clustering}
\DontPrintSemicolon
    Given $\vect{x}_1, \ldots, \vect{x}_n$ points on the Sphere and $M$ number of clusters\\ 
    \nlset{0} Initialize $\Theta$. \\
    \nlset{E}--step:
    \For{k = 1:M}{
    For each observation $\vect{x}_i$, compute the kernel $K_{\rho_k}(\vect{x}| \vect{\mu}_k)$;\\
    Compute the posterior probabilities $p(k| \vect{x}_i, \theta_k) = \frac{\alpha_k f_k(\vect{x}_i|\theta_k)}{f(\vect{x}_i|\Theta)}$;\\
    Compute the weights $w_{ik} = \frac{p(k| \vect{x}_i, \theta_k)}{1 + \rho_k^2 - 2\rho_k \vect{x}_i\vect{\mu}_k}$;\\
    }
    \nlset{M}--step:
    \For{k = 1:M}{
    Estimate parameters as:
    \begin{itemize}
        \item $\alpha_k = 1/n \sum_{i=1}^n p(k| \vect{x}_i, \theta_k)$;
        \item $\mu_k = \frac{\sum_{i=1}^n w_{ik} \vect{x}_i}{||\sum_{i=1}^n w_{ik} \vect{x}_i||}$;
        \item $\rho_k$ is the solution of $g_k(y)=0$ in the interval $(0,1)$, where
            $g_k(y) = \frac{-2ny\alpha_k^{(t-1)}}{1-y^2} + d \left|| \sum_{i=1}^n w_{ik}^{(t-1)} \vect{x}_i\right||  - dy \sum_{i=1}^n w_{ik}^{(t-1)}.$
    \end{itemize}   
    }
    Perform E--M steps until convergence. 
\end{algorithm}
A mixture of $M$ Poisson kernel-based densities given by 
    \begin{equation*}
        f(\vect{x}| \Theta)= \sum_{j=1}^M \alpha_j f_j (\vect{x}| \rho_j, \vect{\mu}_j), 
    \end{equation*}
where $M$ is the number of clusters, $\alpha_j$ are the mixing proportions with $\sum_{j=1}^M \alpha_j =1,$ and $\Theta =\{\alpha_1, \ldots, \alpha_M, \rho_1, \ldots, \rho_M, \vect{\mu}_1, \ldots, \vect{\mu}_M\}$, is the model that is considered.

\subsection{The PKBC class}
In this section, we describe the \texttt{PKBC} class in \texttt{Python} of \texttt{QuadratiK} for clustering data on the sphere. The class \texttt{pkbc} in \texttt{R} has a similar structure, for a detailed description see the package documentation on CRAN. 

An object of the \texttt{PKBC} class can be instantiated using the following parameters:
\begin{itemize}
    \item \texttt{num\_clust}: Number of clusters. It can be a single value or a numeric vector, which can be represented as a list, numpy array, or range.
    \item \texttt{max\_iter}: Maximum number of iterations before a run is terminated. Default is 300.
    \item \texttt{stopping\_rule}: String describing the stopping rule to be used within each run, one of `max', `membership', or `loglik'. Default is `loglik'.
    \item \texttt{init\_method}: String describing the method used for initialization of the centroids. Currently must be 'sampledata'.
    \item \texttt{num\_init}: Number of initializations. Default is 10.
    \item \texttt{tol}: Constant defining threshold by which log likelihood must change to continue iterations, if applicable. Defaults to 1e-7.
    \item \texttt{random\_state}: Determines random number generation for centroid initialization. Defaults to None.
    \item \texttt{n\_jobs}: Used only for computing the WCSS efficiently. \texttt{n\_jobs} specifies the maximum number of concurrently running workers. Default is 4.
\end{itemize}
To execute the PKBD model based clustering algorithm we need to initialize the parameters of the model. The \texttt{sample\_data} method initializes the centroids using randomly chosen observations. Then, the final estimates are those with corresponding highest likelihood. The concentration parameters are initialized at 0.5 and the algorithm starts with equal mixing proportions. The available stopping criteria are: `max': until the change in the log-likelihood is less than a given threshold $(1e-7)$; `membership': until the membership is unchanged; `loglik': based on a maximum number of iterations. The estimation of the number of clusters is an important issue in clustering. Instead of providing a single value for \texttt{num\_clust}, it can be given as a vector of possible number of clusters. For each value in the vector, the function estimates the parameters of the PKBD mixture model which are used for assigning memberships as maximum argument of posterior probabilities. 

The \texttt{PKBC} class contains a \texttt{fit} method to perform the clustering on the data. It takes the data as input, either in the form of a dataframe or a matrix. The \texttt{fit} method returns a fitted class object with the following attributes:

\begin{itemize}
    \item \texttt{alpha\_} Estimated mixing proportions.
    \item \texttt{labels\_} Final cluster membership assigned by the algorithm to each observation.
    \item \texttt{log\_lik\_vecs\_} Vector of log-likelihood values for each initialization.
    \item \texttt{loglik\_} Maximum value of the log-likelihood function.
    \item \texttt{mu\_} Estimated centroids.
    \item \texttt{num\_iter\_per\_runs\_} Number of E-M iterations per run 
    \item \texttt{post\_probs\_} Posterior probabilities of each observation for the indicated clusters.
    \item \texttt{rho\_} Estimated concentration parameters rho.
    \item \texttt{euclidean\_wcss\_} Values of within-cluster sum of squares computed with Euclidean distance.
    \item \texttt{cosine\_wcss\_} Values of within-cluster sum of squares computed with cosine similarity.
\end{itemize}

The attributes are stored as a dictionary, which is a collection of key-value pairs. In this case, each key represents the \texttt{num\_clust}, and the corresponding value is the specific attribute for that particular \texttt{num\_clust}. The \texttt{PKBC} class also provides additional methods that aid in the exploration of the data under investigation. These include methods for predicting the membership of observations using previously estimated cluster centroids (\texttt{predict}), cluster validation (\texttt{validation}), descriptive analysis (\texttt{stats\_clusters}), visualizing data on sphere or circle (\texttt{plot}). Additionally, a \texttt{summary} function is also available to display log-lilelihood, euclidean WCSS, cosine WCSS, cluster sizes and, mixing proportions at a glance. 

In the following section we illustrate the usage of the class \texttt{PKBC} in \texttt{Python}, as well as the supplementary functions, in a real data application. The corresponding codes in \texttt{R} are provided as vignette of the \texttt{QuadratiK} \texttt{R} package and can be accessed via the following link  \href{https://giovsaraceno.github.io/QuadratiK-package/articles/wireless_clustering.html}{https://giovsaraceno.github.io/QuadratiK-package/articles/wireless\_clustering.html}.

\subsection{The Wireless Indoor Localization Data}

We consider the Wireless Indoor Localization Data Set \citep{wireless}, publicly available in the UCI Machine Learning Repository’s website. This data set is used to study the performance of different indoor localization algorithms \citep{rohra2017, wang2023}. It is also available within the \texttt{QuadratiK} package as \texttt{wireless}. Listing \ref{code:wireless} shows the first rows of the \texttt{wireless} data set.

\begin{lstlisting}[language=Python, frame=tb, label=code:wireless, caption=The \texttt{wireless} data set.]
>>> from QuadratiK.datasets import load_wireless_data
>>> wireless_data = load_wireless_data()
>>> print(wireless_data.head())

    WS1   WS2   WS3   WS4   WS5   WS6   WS7  Class
0 -64.0 -56.0 -61.0 -66.0 -71.0 -82.0 -81.0      1
1 -68.0 -57.0 -61.0 -65.0 -71.0 -85.0 -85.0      1
2 -63.0 -60.0 -60.0 -67.0 -76.0 -85.0 -84.0      1
3 -61.0 -60.0 -68.0 -62.0 -77.0 -90.0 -80.0      1
4 -63.0 -65.0 -60.0 -63.0 -77.0 -81.0 -87.0      1

\end{lstlisting}
The Wireless Indoor Localization data set contains the measurements of the Wi-Fi signal strength in different indoor rooms. It consists of a data frame with 2000 rows and 8 columns. The first 7 variables report the values of the Wi-Fi signal strength received from 7 different Wi-Fi routers in an office location in Pittsburgh (USA). The last column indicates the class labels, from 1 to 4, indicating the different rooms. Notice that, the Wi-Fi signal strength is measured in dBm, decibel milliwatts, which is expressed as a negative value ranging from -100 to 0. In total, we have 500 observations for each room.
Given that the Wi-Fi signal strength takes values in a limited range, it is appropriate to consider the spherically transformed observations, by $L_2$ normalization, and consequently perform the clustering algorithm on the 7-dimensional sphere.

\begin{landscape}
\begin{figure}[htbp]
    \centering
    \includegraphics[width=0.9\linewidth]{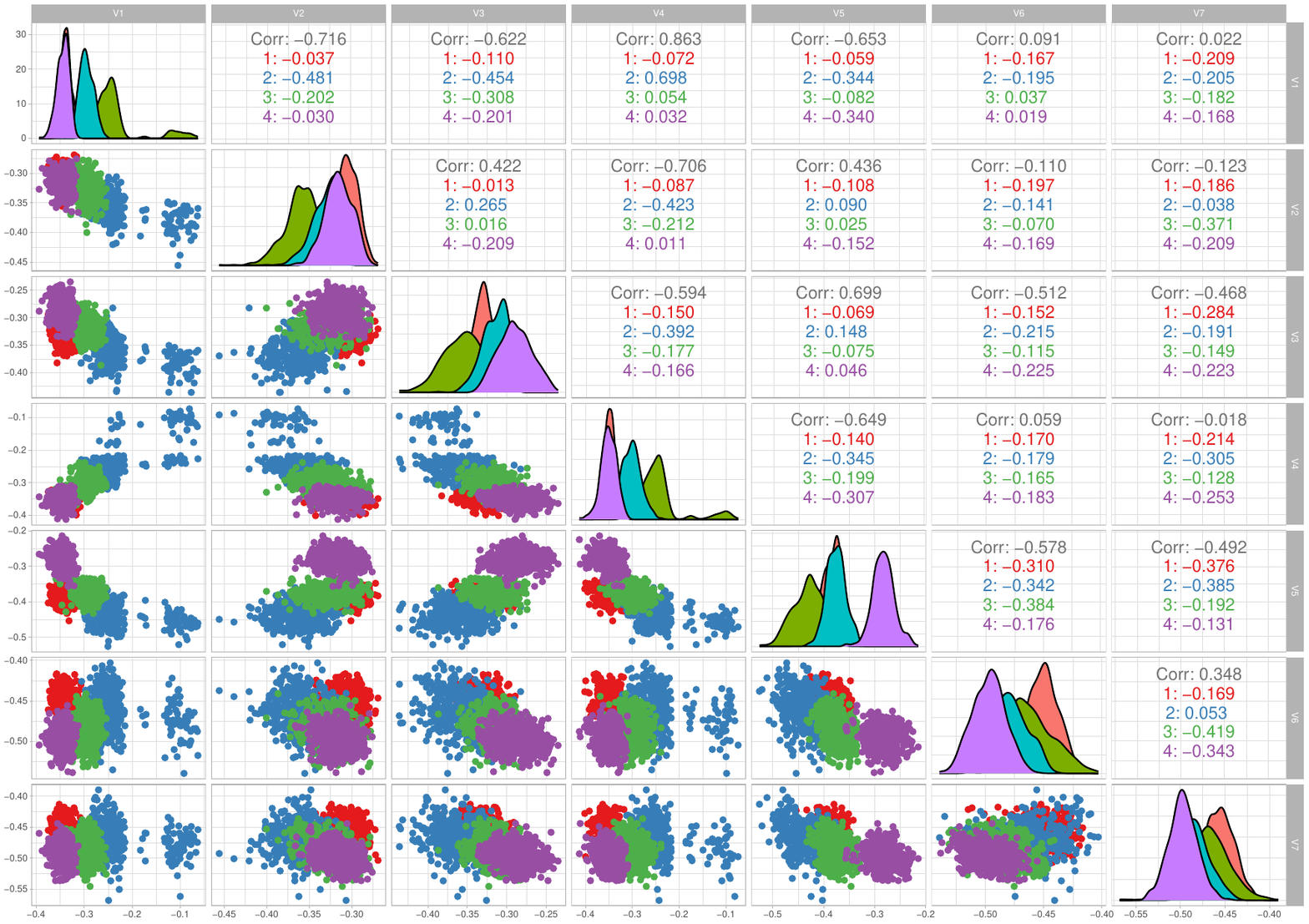}
    \caption{Set of plots of the \texttt{wireless} data with spherically transformed data points, displayed with respect to the given labels. Density plots of individual variables (main diagonal); pair-wise scatter plots (lower triangular); pair-wise correlations, overall and per group.}
    \label{fig:wireless_pairs_norm}
\end{figure}
\end{landscape}

Figure \ref{fig:wireless_pairs_norm} shows a set of plots displayed with respect to the given labels of the \texttt{wireless} data set after that data points have been normalized, and it is generated using the function \texttt{ggpairs} in the \texttt{R} package \texttt{GGally} \citep{GGally}. The figure displays the density plot of individual variables and pair-wise scatter plots colored by the given labels. Additionally, the boxes in the upper triangular matrix show the pair-wise correlations, overall and by group. 
We have verified that the overall structure in the data set is preserved with normalization. 

We perform the clustering algorithm on the \texttt{wireless} data set. Listing \ref{code:wireless-pkbc} shows the \text{R} codes which illustrates the usage of the \texttt{pkbc} function. We consider the $K= 2, \ldots, 10$ as possible values for the number of clusters. 
\begin{lstlisting}[language=Python, frame=tb, label=code:wireless-pkbc, caption=Spherical clustering of the \texttt{wireless} data using the \texttt{PKBC} and its \texttt{fit} method]
>>> from QuadratiK.datasets import load_wireless_data
>>> from QuadratiK.spherical_clustering import PKBC
>>> X, y = load_wireless_data(return_X_y=True)
>>> pkbc = PKBC(num_clust=range(2,11), random_state=42).fit(X)
\end{lstlisting}

To guide the choice of the number of clusters, we provide a function which computes cluster validation measures and includes graphical tools. Specifically \texttt{pkbc\_validation}, in \texttt{R}, returns an object with \texttt{IGP}, the InGroup Proportion (IGP) \citep{Kapp2006}, and \texttt{metrics} a table of computed evaluation measures. This includes the Average Silhouette Width (ASW) \citep{ROUSSEEUW198753} and, if the true labels are provided, the measures of adjusted Rand index (ARI), Macro-Precision and Macro-Recall.
Similarly, using \texttt{validation} method in \texttt{Python}, a table with evaluation measures and elbow plots are returned.

\begin{figure}[htbp]
    \centering
    \includegraphics[width=\linewidth]{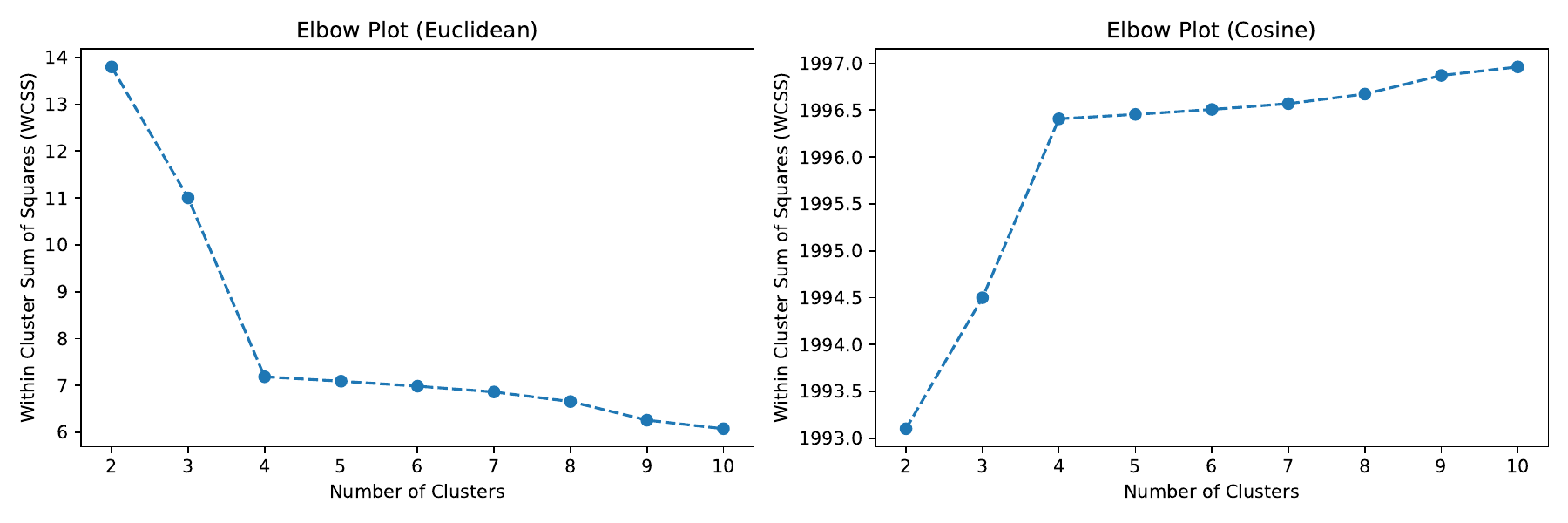}
    \caption{Elbow plots of the within-cluster sum of squares, using the Euclidean distance (left) and cosine similarity (right), versus the number of clusters, from 2 to 10, for the \texttt{wireless} data.}
    \label{fig:wireless_elbow}
\end{figure}

Listing \ref{code:wireless-validation} illustrates the usage of the \texttt{validation} method on the clustering results of the \texttt{wireless} data set, together with the obtained results. Figure \ref{fig:wireless_elbow} shows the generated elbow plots. 
\begin{lstlisting}[language=Python, frame=tb, label=code:wireless-validation, caption=The \texttt{validation} function for the \texttt{wireless} data set., 
basicstyle=\footnotesize\ttfamily]
>>> import pandas as pd
>>> validation_metrics, elbow_plots = pkbc.validation(y_true = y)
>>> print(validation_metrics.round(2))
>>> elbow_plots

                             2     3     4     5     6     7     8     9    10
Metrics                                                                       
ARI                       0.31  0.70  0.94  0.91  0.88  0.80  0.73  0.69  0.60
Macro Precision           0.31  0.61  0.98  0.98  0.98  0.98  0.97  0.98  0.97
Macro Recall              0.50  0.75  0.98  0.98  0.98  0.98  0.97  0.98  0.97
Average Silhouette Score  0.42  0.35  0.38  0.30  0.20  0.13  0.11  0.08  0.10

\end{lstlisting}

\begin{landscape}
\begin{figure}[htbp]
    \centering
    \includegraphics[width=\linewidth]{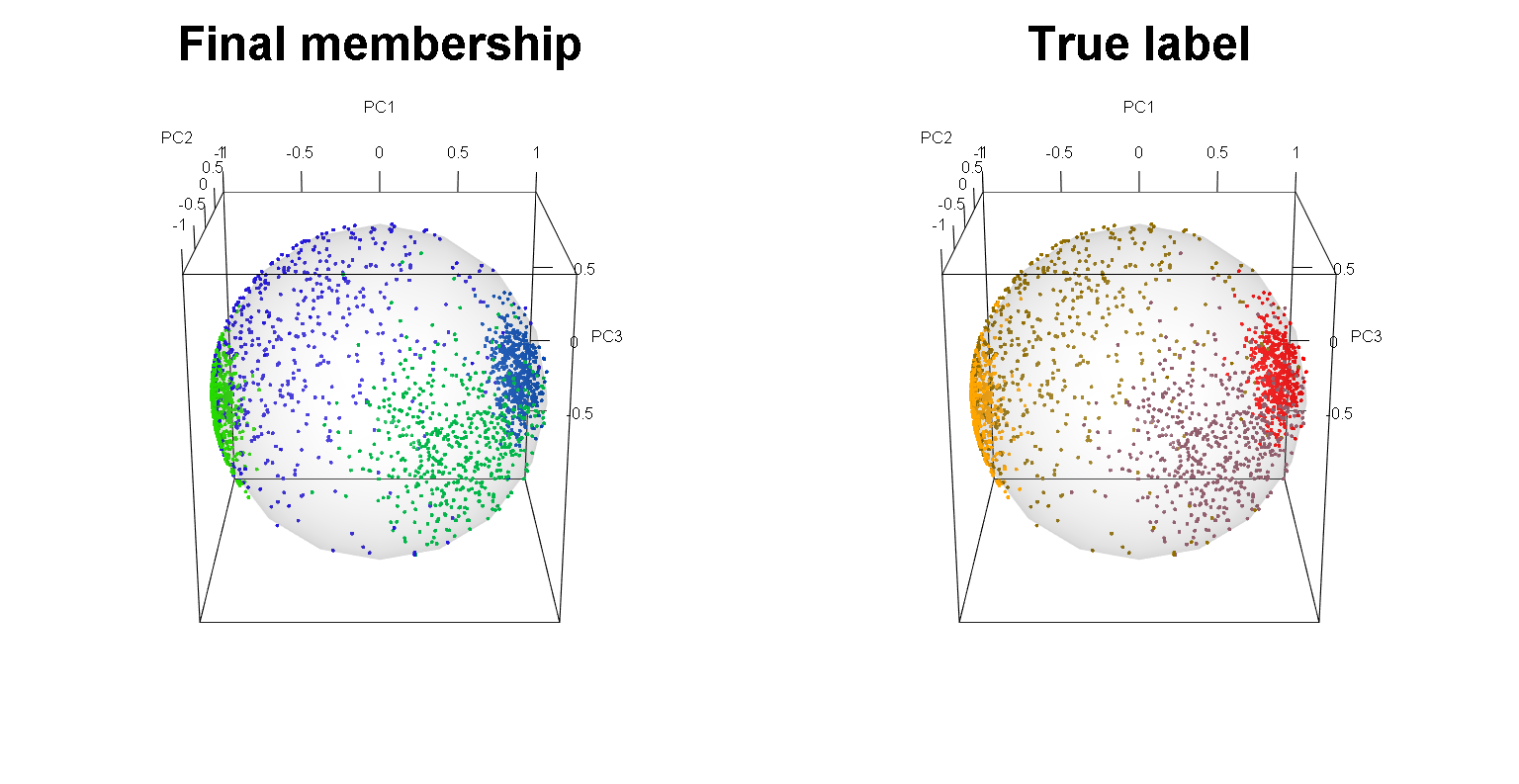}
    \caption{Data points of the \texttt{wireless} data set displayed according to the first three principal components from spherical PCA, after applying $L_2$ normalization in \texttt{R}. Observations are colored with respect to the final memberships assigned by the clustering algorithm for $K=4$ clusters (left) and the true label (right).}
    \label{fig:wireless_sphere}
\end{figure}
\end{landscape}

The elbow plots and the reported metrics suggest $K=4$ as number of clusters. This is in accordance with the known ground truth. 
Once the number of clusters is selected, the function \texttt{stats\_clusters} in the \texttt{QuadratiK} package can be used to obtain additional summary information with respect to the clustering results. In particular, the function provides mean, standard deviation, median, inter-quantile range, minimum and maximum computed for each variable, overall and by the assigned membership. In both \texttt{Python} and \texttt{R} implementation, we also provide a \texttt{plot} method that can be utilized to generate a graphical representation of data points. For $d=2$ and $d=3$, observations are displayed on the unit circle and unit sphere, respectively. 
In the \texttt{Python} implementation, if $d>3$, the PCA is applied to the data set, and the first 3 principal components are used to visualize the data on the sphere, after applying $L_2$ normalization. Meanwhile in the \texttt{R} implementation, if $d>3$, the spherical PCA is applied on the data set, and the first 3 principal components are used for visualizing data points on the sphere, after application of  $L_2$ normalization. Listing \ref{code:wireless-summary} shows the usage of the \texttt{stats\_clusters} function in \texttt{Python} on the clustering results of the \texttt{wireless} data set. Figure \ref{fig:wireless_sphere} is generated in \texttt{R} and shows the data points according to the first three spherical principal components, colored by the true labels and the assigned membership. 
\begin{lstlisting}[language=Python, frame=tb, label=code:wireless-summary, caption=The \texttt{stats\_clusters} function for the \texttt{wireless} data set., basicstyle=\footnotesize\ttfamily]
>>> print(pkbc.stats_clusters(num_clust = 4))

                     Group 0    Group 1    Group 2    Group 3    Overall
Feature 0 Mean    -60.180000 -36.740741 -62.380952 -49.558824 -52.330500
          Std Dev   3.025543   8.768657   3.478093   2.985142  11.321677
          Median  -60.000000 -38.000000 -62.000000 -50.000000 -55.000000
          IQR       4.000000   5.000000   5.000000   4.000000  15.000000
          Min     -71.000000 -52.000000 -74.000000 -62.000000 -74.000000
          Max     -52.000000 -10.000000 -48.000000 -38.000000 -10.000000
Feature 1 Mean    -55.208000 -56.181070 -56.240079 -54.890196 -55.623500
          Std Dev   3.251135   3.346660   3.288014   3.574042   3.417688
          Median  -55.000000 -56.000000 -56.000000 -55.000000 -56.000000
          IQR       4.000000   4.000000   4.000000   4.750000   5.000000
          Min     -66.000000 -74.000000 -69.000000 -68.000000 -74.000000
          Max     -46.000000 -46.000000 -47.000000 -45.000000 -45.000000
Feature 2 Mean    -50.652000 -55.989712 -60.416667 -52.825490 -54.964000
          Std Dev   4.144061   4.198416   3.812322   3.115797   5.316186
          Median  -50.500000 -56.000000 -60.000000 -52.500000 -55.000000
          IQR       6.000000   5.000000   5.000000   4.000000   7.000000
          Min     -60.000000 -71.000000 -73.000000 -73.000000 -73.000000
          Max     -40.000000 -46.000000 -47.000000 -44.000000 -40.000000
Feature 3 Mean    -61.310000 -37.792181 -64.109127 -50.588235 -53.566500
          Std Dev   3.857936   7.915274   3.737670   3.651716  11.471982
          Median  -61.000000 -39.000000 -64.000000 -51.000000 -56.000000
          IQR       5.000000   6.000000   4.500000   4.000000  17.000000
          Min     -77.000000 -51.000000 -77.000000 -61.000000 -77.000000
          Max     -52.000000 -11.000000 -51.000000 -37.000000 -11.000000
Feature 4 Mean    -49.426000 -67.806584 -70.202381 -63.200000 -62.640500
          Std Dev   3.505582   5.224411   4.664352   3.475313   9.105093
          Median  -50.000000 -68.000000 -69.000000 -63.000000 -64.000000
          IQR       5.000000   7.000000   7.000000   5.000000  13.000000
          Min     -61.000000 -86.000000 -89.000000 -77.000000 -89.000000
          Max     -36.000000 -56.000000 -60.000000 -54.000000 -36.000000
Feature 5 Mean    -87.012000 -72.512346 -82.823413 -81.333333 -80.985000
          Std Dev   3.393950   4.573664   3.747838   3.739381   6.516672
          Median  -87.000000 -72.000000 -82.000000 -80.000000 -82.000000
          IQR       4.000000   6.000000   5.000000   5.000000   9.000000
          Min     -96.000000 -89.000000 -97.000000 -93.000000 -97.000000
          Max     -76.000000 -61.000000 -74.000000 -71.000000 -61.000000
Feature 6 Mean    -86.986000 -73.341564 -83.922619 -82.390196 -81.726500
          Std Dev   3.546118   4.655847   3.980065   4.334118   6.519812
          Median  -87.000000 -73.000000 -83.000000 -82.000000 -83.000000
          IQR       4.000000   7.000000   6.000000   7.000000   9.000000
          Min     -98.000000 -90.000000 -96.000000 -93.000000 -98.000000
          Max     -78.000000 -63.000000 -74.000000 -69.000000 -63.000000
\end{lstlisting}
The clusters identified with $k=4$ achieve high performance metrics in terms of ARI, Macro Precision and Macro Recall. Figure \ref{fig:wireless_sphere} also shows that the identified clusters follows the initial labels.

\section{Summary}
\label{sec:conclusion}
In this article, we presented a new, innovative package that implements goodness of fit testing and clustering methods that depend on kernel-based quadratic distance technologies. The package, named \texttt{QuadratiK}, incorporates many functionalities, such as testing goodness of fit for one, two and k-sample problems, uniformity testing on the $d$-dimensional sphere and an algorithm for clustering data on the $d$-dimensional sphere.

The implemented methodology has several advantages, not withstanding the fact that the methods are mathematically sound and the software is easy to use. Its implementation in both \text{R} and \text{Python} programming languages aims to attract a wider group of users.

\section*{Acknowledgments}

We acknowledge help provided by Dr. Yang Chen, Associate Director, GI Statistics, Takeda Pharmaceutical Company. The second author would like to acknowledge financial support in the form of a research award from KALEIDA Health Foundation and FDA, that funded the work of the first and third authors.  

\bibliography{kernel_ACM.bib}

\end{document}